\documentclass[fleqn,usenatbib,useAMS]{mnras}

\usepackage{savesym}
\usepackage{graphicx}	
\usepackage{amsmath}	
\usepackage{amssymb}	
\usepackage{multicol}        
\usepackage{bm}		
\usepackage{pdflscape}	

\usepackage{verbatim}
\savesymbol{comment}
\usepackage[final, markup=bfit, deletedmarkup=sout, authormarkup=none]{changes}
\restoresymbol{ch}{comment}

\usepackage[T1]{fontenc}
\usepackage{ae,aecompl}

\usepackage{newtxtext,newtxmath}
\definechangesauthor[color=red]{AAB}

\title[Q0107: Gas around Galaxies]{The relationship between gas and galaxies at $z<1$ using the Q0107 quasar triplet}

\author[Beckett et al.]{Alexander Beckett$^{1}$\thanks{Contact e-mail: \href{mailto:alexander.beckett@durham.ac.uk}{alexander.beckett@durham.ac.uk}},
Simon L. Morris$^{1}$,
Michele Fumagalli$^{3,2,1}$,
Rich Bielby$^{1}$,
\newauthor
Nicolas Tejos$^{4}$, Joop Schaye$^{5}$, Buell Jannuzi$^{6}$, Sebastiano Cantalupo$^{3, 7}$
\\
$^{1}$Centre for Extragalactic Astronomy, Durham University, South Road, Durham DH1 3LE, UK\\
$^{2}$Institute for Computational Cosmology, Durham University, South Road, Durham, DH1 3LE, UK\\
$^{3}$Dipartimento di Fisica G. Occhialini, Universit\`{a} degli Studi di Milano Bicocca, Piazza della Scienza 3, 20126 Milano, Italy\\
$^{4}$Instituto de F\'{i}sica, Pontificia Universidad Cat\'{o}lica de Valpara\'{i}so, Casilla 4059, Valpara\'{i}so, Chile\\
$^{5}$Leiden Observatory, Leiden University, PO Box 9513, 2300, RA Leiden, The Netherlands\\
$^{6}$Steward Observatory, University of Arizona, 933 North Cherry Avenue, Tucson, AZ 85721, USA\\
$^{7}$Department of Physics, ETH Zurich, Wolfgang-Pauli Strasse 27, Zurich, Switzerland
}
\date{Last updated 20XX Jan 1; in original form 20XX Jan 1}

\pubyear{2020}

\begin{document}
\label{firstpage}
\pagerange{\pageref{firstpage}--\pageref{lastpage}}
\maketitle

\begin{abstract}
We study the distribution and dynamics of the circum- and intergalactic medium using a dense galaxy survey covering the field around the Q0107 system, a unique \hbox{z $\approx$ 1} projected quasar triplet. With full Ly$\alpha$ coverage along all three lines-of-sight from 
z=0.18 to z=0.73, more than 1200 
galaxy spectra, and two MUSE fields, we examine the structure of the gas around galaxies on 100-1000 kpc scales. We search for \ion{H}{i} absorption systems occurring at the same redshift \added[id=AAB]{(within 500 $\textrm{km}$ $\textrm{s}^{-1}$)} in multiple sightlines, finding with \hbox{$>$ 99.9\%} significance that these systems are more frequent in the observed quasar spectra than in a randomly distributed population of absorbers. This is driven primarily by absorption with column densities N(\ion{H}{i}) $> 10^{14}$ $\textrm{cm}^{-2}$, whilst multi-sightline absorbers with lower column densities are consistent with a random distribution. Star-forming galaxies are more likely to be associated with multi-sightline absorption than quiescent galaxies. \added[id=AAB]{HST imaging provides inclinations and position angles for a subset of these galaxies.} We observe a bimodality in the position angle of detected galaxy-absorber pairs, again driven mostly by high-column-density absorbers, with absorption preferentially along the major and minor axes of galaxies out to impact parameters of several hundred kpc. We find some evidence supporting a disk/outflow dichotomy, as \ion{H}{i} absorbers near the projected major-axis of a galaxy show line-of-sight velocities that tend to align with the rotation of that galaxy, whilst minor-axis absorbers are twice as likely to exhibit \ion{O}{vi} at the same redshift.




\end{abstract}


\begin{keywords}
intergalactic medium -- quasars: absorption lines -- galaxies: formation -- large-scale structure of Universe
\end{keywords}



\newpage

\section{Introduction}

Galaxies follow a large-scale filamentary structure throughout the Universe, known as the cosmic web \citep{bond1996}, formed by the gravitational accretion of gas towards the potential wells of dark matter around initial overdensities, as has been modelled in dark-matter-only simulations for decades \citep[e.g.][]{klypin1983, springel2005}. Once stars and galaxies form, this accretion from the intergalactic medium (IGM) towards galaxies becomes affected by complex baryonic physics, including stellar and AGN feedback \citep[e.g.][]{vandevoort2011, nelson2019, mitchell2020}. Simulating these effects on a range of scales from sub-parsec-scale supernovae to megaparsec-scale gas flows around clusters necessitates sub-grid models which require constraints from observations. 

Such constraints are necessary for studies of galaxy evolution, as exchanges of material between galaxies and their environments play an important role in regulating star formation \added[id=AAB]{\citep[e.g.][]{keres2005, schaye2010, dave2012, lehnert2013, somerville2015, salcido2020}}. Gas outside of galaxies is believed to contain a substantial fraction of the baryons in the Universe \citep[e.g][]{fukugita1998, behroozi2010a, werk2014}, so any census of baryons used to constrain cosmological parameters must consider the state of the gas around galaxies (usually by modelling the ionization state of the gas based on absorption spectra, e.g. \citealt{shull2012}, although recent methods using fast radio bursts account for all ionized baryons, e.g. \citealt{macquart2020}). The dynamics of galaxies are also strongly linked to the state of the surrounding gas through the transfer of angular momentum \added[id=AAB]{\citep[e.g][]{pichon2011, stewart2017, defelippis2020}}. Therefore, observational insights into the distribution of baryons around galaxies can not only inform our understanding of the gas itself, but also stellar processes within galaxies and their effects on galaxy formation and evolution.

Outside the local environment, most observations of the gas around galaxies are made by identifying absorption features in the spectrum of background sources, usually quasars. These features can probe neutral gas, most commonly using \ion{H}{i} \added[id=AAB]{\citep[e.g.][]{morris1991, lanzetta1995, chen1998, adelberger2003, tumlinson2013, rakic2013, heckman2017, chen2018}}, \added[id=AAB]{especially since the Hubble Space Telescope has allowed the Lyman-$\alpha$ transition to be observed at low redshifts. Other studies} use low ions such as \ion{Mg}{ii} to probe cool ($\sim 10^{4} \textrm{K}$) gas \added[id=AAB]{\citep[e.g.][]{bergeron1986, bergeron1991, bouche2006, nielsen2013, schroetter2016, ho2017}}, or search for highly-ionized material \added[id=AAB]{\citep[e.g.][]{bergeron1994, tripp2000, cen2001, tumlinson2011, turner2014, finn2016, werk2016, nicastro2018, bielby2019}} through transitions from ions such as \ion{O}{vi} to \ion{O}{viii}. This variety of observed ions found in sightlines probing the gas near to galaxies suggests that it has a complex, multi-phase structure \added[id=AAB]{\citep[e.g.][]{veilleux2005, werk2013, mathes2014, peroux2019, chen2020a}}.

Galaxy-scale outflows are observed in emission \citep[e.g.][]{bland1988, finley2017, burchett2021} and absorption \citep[e.g.][]{grimes2009, turner2015, lan2018, schroetter2019}, their multi-phase nature indicated by a range of diagnostics from low ions (tracing cool gas) \citep[e.g.][]{concas2019}, to X-ray-emitting hot gas \citep[e.g.][]{lehnert1999}. These are consistent with stellar-feedback, in which supernovae and stellar winds drive material out of the galaxy \hbox{\citep[e.g.][]{chevalier1985, heckman1990}} to distances of several tens of kiloparsecs. Cosmological simulations also find these winds, and can produce the multi-phase biconical outflows observed despite isotropic injection of energy and momentum \citep[e.g.][]{nelson2019, mitchell2020a}. \citet{hopkins2021} also find that the addition of cosmic rays to the outflow-driving mechanism may allow such flows to reach megaparsec scales.


\added[id=AAB]{There are several lines of evidence suggesting that substantial gas accretion occurs onto galaxies from the surrounding medium, including the metallicities of dwarf stars \citep[e.g.][]{casuso2004}, the short depletion timescales of star-forming galaxies \citep[e.g.][]{freundlich2013, scoville2017}, and the declining \ion{H}{i} density of the Universe over time \citep[e.g.][]{neeleman2016}.}
Evidence for this accretion is found in `down-the-barrel' observations (using the host galaxy as the background source) where absorption line profiles indicate gas flows towards the galaxy \citep[e.g][]{martin2012, rubin2012}. Observations \added[id=AAB]{in \ion{Mg}{ii} have long suggested that the gas shows strong rotation, often with an in-falling velocity component \citep[e.g][]{charlton1998, steidel2002}. When this is compared to the rotation curves of galaxies} , often using integral field units such as MUSE, \added[id=AAB]{most} find that absorbing gas close to the major axis of a galaxy preferentially shows co-rotation in both \ion{Mg}{ii} \citep[e.g.][]{ho2017, martin2019a, zabl2019} and \ion{H}{i} \citep[e.g][]{french2020}, particularly within 100 kpc. 
\added[id=AAB]{However,} this co-rotation is not always apparent when extending to weaker absorbers or larger impact parameters \citep[e.g.][]{dutta2020}.

This `galactic fountain' model, consisting of minor axis outflows and major axis co-rotating accretion, can explain the bimodality in position angle found in the MEGAFLOW survey \citep{schroetter2016, zabl2019, schroetter2019}. It is also supported by cosmological simulations, for example the FIRE simulations \citep{hafen2019a, hafen2020} produce galaxies around which the gas around galaxies is a mixture of material ejected from the galaxy interstellar medium (ISM) in winds, and material accreted from the IGM. Much of this material then accretes onto the central galaxy. However, this model would predict generally lower metallicities along the major axis, which has not been detected in recent \ion{H}{i} observations \citep{pointon2019, kacprzak2019}, and \ion{Mg}{ii} observations often do not find a significant bimodality in position angle \citep{dutta2020, huang2021}. How frequently these structures form, and to what extent, therefore remains uncertain.


Whilst a galactic fountain can reproduce many observations of the gas close to galaxies (the circumgalactic medium, or CGM; see \citealt{tumlinson2017} for a recent review), studies of the correlation between gas and galaxies show transverse correlation lengths of 2--3 Mpc at z $<$ 1 \citep{tejos2014, finn2016}. On these scales the interactions between galaxies are expected to have a significant effect on the structure of the CGM/IGM \citep[e.g][]{fossati2019, dutta2020}. Intra-group material can be observed in absorption, that can not easily be assigned to an individual galaxy \citep[e.g.][]{peroux2017, bielby2017a, chen2020a}. We expect interactions in groups and clusters to build up pressure-confined tidal debris, visible across a range of column densities \citep{morris1994}. This is clearly visible in \ion{H}{i} maps of the Magellanic Stream \citep[e.g.][]{mathewson1974, nidever2008} and the M81/M82 group \citep[e.g.][]{croxall2009, sorgho2019}, as well as H$\alpha$ and occasionally [\ion{O}{iii}] emission in dense environments \citep[e.g][]{fumagalli2014, johnson2018, fossati2019a}.

Absorption, especially in Ly$\alpha$, is our most sensitive probe of the CGM and IGM, but it is in most cases limited to a single pencil-beam sightline through the CGM of any individual galaxy. This makes it difficult to directly constrain the size scales and structures in the CGM, although some studies have stacked large samples together attempting to infer these properties \citep[e.g][]{chen2012, turner2014}.

Additional information can be extracted using multiple sightlines probing the CGM of an individual galaxy. \citet{bowen2016} take advantage of the large angle subtended by relatively nearby haloes, such that the halo is pierced by a number of QSO sightlines, finding the gas surrounding NGC 1097 to have a disk-like structure with rotating and in-falling velocity components. \added[id=AAB]{Similarly \citet{keeney2013} find the absorption in three sightlines around a nearby galaxy to be consistent with a `galactic fountain'.} Gravitational lensing produces multi-sightline systems, through multiple images \added[id=AAB]{\citep[e.g][]{smette1992, rauch2002, chen2014, zahedy2016}}, and extended arcs that can be considered as multiple closely-spaced lines-of-sight \added[id=AAB]{\citep[e.g.][]{lopez2018, lopez2020}}. It is also possible to use bright background galaxies as sources (which may also be extended), as an alternative or in addition to quasar sightlines \added[id=AAB]{\citep[e.g.][]{adelberger2005, steidel2010, cooke2015, rubin2018, fossati2019, zabl2020}}. 

This study focuses on the Q0107 system, a quasar triplet at z $\approx$ 1: LBQS 0107-025A, LBQS 0107-025B, and LBQS 0107-0232, hereafter denoted A, B, and C. Table \ref{table:past_papers} summarizes some of the main parameters of this system. This allows multiple sightlines to be probed through the CGM/IGM around galaxies in this field, separated by hundreds of kpc.

\begin{table}\label{table:past_papers}
\begin{center}
\caption{Co-ordinates, redshifts and R-band magnitudes of the three quasars, taken from \citet{crighton2010}}
\begin{tabular}{| c| c| c| c| c|}
 \hline Object & RA (J2000) & Dec (J2000) & Redshift & R-mag \\ \hline
 Q0107-025 A & 01:10:13.14 & -2:19:52.9 & 0.960  & 18.1\\
 Q0107-025 B & 01:10:16.25  & -2:18:51.0 & 0.956 & 17.4\\
 Q0107-0232 (C) & 01:10:14.43 & -2:16:57.6 & 0.726 & 18.4\\ \hline

\end{tabular}

\end{center}
\end{table}

As the only known, bright, low-redshift quasar triplet, this system has been the focus of many studies. \citet{dinshaw1997} observed A and B, finding 5 absorption systems that cover both sightlines, and 6 limited to a single sightline. Using a maximum likelihood analysis, they concluded that their data was best explained by randomly inclined disks approximately 1 Mpc in radius. \citet{young2001} complemented this with analysis of the likelihood of individual multi-sightline absorption systems, and found that matches involving stronger absorption features tended to have smaller velocity separations. Coincidences of absorption between the sightlines also occur more frequently among high-column-density absorbers, as found by \citet{petry2006}.

A later study by \citet{crighton2010} (hereafter C10) used improved QSO spectra, including QSO-C, in addition to galaxy data from CFHT-MOS, to extend these results. They observed a highly-significant excess of absorption systems covering all three sightlines over an ensemble of randoms, providing clear evidence that gas and galaxies are associated on scales of hundreds of kpc. Additionally, galaxies and groups of galaxies could be associated with multiple absorbers, allowing the structure of the gas to be analysed in individual systems (although we defer an updated analysis of these systems to a later paper, focusing here on the statistical properties of our samples).

Ionization modelling was used by \citet{muzahid2014} to study one example in this field at z $\sim$ 0.22, using the presence of \ion{O}{vi} in sightlines A and B to estimate the radius of the CGM as 330 kpc, and therefore detect both the warm and cool CGM of an $L_{\star}$ galaxy.

This field was also included in a study of galaxy--absorber cross-correlations covering six independent fields by \hbox{\citet{tejos2014}} (hereafter T14) and another using 50 fields by \citet{finn2016} (F16). Their galaxy catalogue forms the basis for the galaxy data used in this work.
 
In this paper we present an updated analysis on the Q0107 triplet, using a much larger sample of galaxies extending to fainter magnitudes, in addition to Hubble Space Telescope imaging providing improved morphologies, and MUSE fields providing kinematics on a subsample of galaxies close to the A and B sightlines. We use this data to examine the CGM/IGM on large scales, where the improved imaging allows us to constrain the extent of the `galactic fountain' and the larger galaxy samples allow us to study the presence of absorption features covering multiple sightlines around galaxies of different properties.

In Section \ref{sec:data} we describe the quasar spectra and galaxy survey used to produce our catalogues of absorption features and galaxies. Section \ref{sec:coherence} discusses our test for correlated absorption between the three sightlines. Section \ref{sec:orientations} gives results from studying the relationship between absorption properties and the position angles of nearby galaxies, whilst in Section \ref{sec:kinematics} we use the MUSE data to identify co- and counter-rotation among material close to the major axis of galaxies. In Section \ref{sec:conclusions} we discuss the consequences of our results and how future work can further progress our understanding.


We use the cosmology from \citet{planckcollaboration2020} throughout, with $\Omega_{m}$ = 0.315 and $H_{0}$ = 67.4 $\textrm{km}$ $\textrm{s}^{-1}$ $\textrm{Mpc}^{-1}$, and quote physical sizes and distances unless otherwise stated.

\section{Data}\label{sec:data}


Our dataset consists of HST/COS and FOS spectra of the three quasars and galaxy surveys from the VIMOS, DEIMOS, GMOS and CFHT-MOS instruments, supplemented by HST R-band imaging and two MUSE fields. In this section we describe the reduction processes for each of these observations and the compilation of the results into our final catalogues of galaxies and absorbers, which are then used in Sections \ref{sec:coherence} - \ref{sec:kinematics}.

\subsection{IGM data}\label{sec:igm_data}

The UV spectra of the quasars were taken by the Cosmic Origins Spectrograph \citep[COS, ][]{green2012} and Faint Object Spectrograph (FOS) on the Hubble Space Telescope . The COS spectra were observed in 2010-11 (program G011585, PI: Neil Crighton), using the G130M and G160M gratings, with a FWHM of 0.07--0.09\AA, and a signal-to-noise ratio (SNR) per pixel of 7--9. These complement longer-wavelength FOS data described in \citet{young2001}. The observations are detailed in Table \ref{table:qso_summary}. The G130M and G270H gratings were not used for QSO-C, due to the Lyman Limit of a sub-damped Lyman$\alpha$ system in the sightline and the lower redshift of this quasar respectively. These observations cover Ly$\alpha$ in COS from $z = 0$ (or the Lyman Limit in the case of QSO-C) to $z \approx 0.45$, and in FOS from 0.45 to the redshift of the quasar.

\begin{table*}
\begin{center}
\caption{Summary of the QSO spectra used to generate the absorption-line catalogue. Columns are: (1) The QSO featured in the spectrum, (2) the spectrograph used, (3) the grating used, (4) the wavelength range for which this instrument and grating provides the spectrum, (5) the full width at half-maximum of the line-spread function of the spectrograph, (6) dispersion of the spectrum, (7) average signal-to-noise across the wavelength range given, (8) exposure time of the observations, (9) HST program ID of the observations. }
\label{table:qso_summary}
\begin{tabular}{| c| c| c| c| c| c| r| r| c|}
\hline 
QSO & Instrument & Grating & Wavelength Range & FWHM & Dispersion & SNR & Exposure Time & Prog ID \\

 & & & (\AA) & (\AA) & (\AA/pix) & (per pix) & (h) & \\
 (1) & (2) & (3) & (4) & (5) & (6) & (7) & (8) & (9) \\ \hline

Q0107-025A & COS & G130M & 1135-1460 & 0.07 & 0.01 & 9 & 7.8 & 11585 \\
 & COS & G160M & 1460-1795 & 0.09 & 0.01 & 8 & 12.3 & 11585 \\
 & FOS & G190H & 1795-2310 & 1.39 & 0.36 & 28 & 7.5 & 5320, 6592 \\
 & FOS & G270H & 2310-3277 & 1.97 & 0.51 & 32 & 2.4 & 6100 \\
 Q0107-025B & COS & G130M & 1135-1460 & 0.07 & 0.01 & 9 & 5.9 & 11585 \\
  & COS & G160M & 1460-1795 & 0.09 & 0.01 & 7 & 5.9 & 11585 \\
  & FOS & G190H & 1795-2310 & 1.39 & 0.36 & 28 & 1.8 & 5320, 6592 \\
  & FOS & G270H & 2310-3277 & 1.97 & 0.51 & 32 & 1.8 & 6100 \\
 Q0107-0232 (C) & COS & G160M & 1434-1795 & 0.09 & 0.01 & 7 & 23.2 & 11585 \\
  & FOS & G190H & 1795-2310 & 1.39 & 0.36 & 18 & 9.1 & 11585 \\
\hline

\end{tabular}

\end{center}
\end{table*}

We use the line catalogue from T14 (also used by F16). They provide a more detailed description of the reduction process with further references, but we summarise it here. Individual exposures from COS were downloaded from the Space Telescope Science Institute (STScI) archive and reduced using CALCOS v2.18.5. 


T14 performed their own background smoothing procedure masking out portions of the spectra affected by strong geocoronal emission lines (namely the Ly$\alpha$ and  \ion{O}{i} 1302,1306 {\AA} lines) and pixels with bad data quality flags. The error array was calculated in the same way as in CALCOS, but using the new background estimation, interpolated across the masked-out regions. Each spectrum was then flux calibrated using sensitivity curves provided by STScI.

Co-alignment was performed by cross-correlating strong galactic absorption features. Pixels with bad data quality flags were then excluded, whilst pixels with warning flags were halved in weight, before fluxes were re-binned to have a constant spacing equal to the dispersion of the grating. Co-addition was weighted by exposure time, and was followed by re-binning on a linear scale sufficiently narrow to ensure Nyquist sampling\footnote{Usually 2 pixels per resolution element, although this is not well-defined for the non-Gaussian line-spread function of COS. Here the FWHM is used to estimate the required sampling.} across the entire wavelength range (0.0395 {\AA} per pixel). 

Individual exposures from FOS were downloaded from the STScI archive and reduced using the standard CALFOS pipeline. Wavelength corrections given by \citet{petry2006} were applied to each individual exposure. The shortest wavelength region of the FOS G190H settings overlap with the longest wavelength COS settings, and T14 confirmed that the wavelength scales in these overlapping regions were consistent between the two instruments. All individual exposures were then combined together, resampling to a common wavelength scale of 0.51 {\AA} per pixel.


T14 then estimated the continuum of each spectrum using a semi-automated method. They split the spectrum into `chunks' of $\approx 12$ {\AA} (blueward of QSO Ly$\alpha$ emission, longer chunks at longer wavelengths), fit a straight line through the points within each chunk (iteratively removing outliers until convergence), then fit a cubic spline to give a smooth result. They checked the resulting continuum `by eye' to ensure a reasonable fit (see their Figure 1).

VPFIT \citep{carswell2014} was used to estimate redshifts, column densities and Doppler parameters for each identifiable absorption system. The \ion{H}{i} systems were assigned a flag (`a', `b' or `c') based on the number of absorption lines observed and the signal-to-noise in the column density estimate:

\begin{description}
    \item (a): At least two Lyman transitions observed with $\textrm{log(N)}$ estimates at least 30 times their uncertainty
    \item (b): Only Ly$\alpha$ observed with $\textrm{log(N)}$ estimate at least 30 times its uncertainty
    \item (c): $\textrm{log(N)}$ estimate less than 30 times its uncertainty
\end{description}

Only those with `a' and `b' flags are included in our analysis. The distribution of \ion{H}{i} column densities is consistent (using results from \citealt{keeney2012}, as discussed in Figure 5 and Section 4.5 of T14) with a 3$\sigma$ detection limit estimate of $\sim10^{13} \textrm{cm}^{-2}$ in the COS spectra and $\sim10^{13.5} \textrm{cm}^{-2}$ in the FOS spectra. Our catalogue should therefore be complete above this column density, with the exception of unresolved blended systems more likely to be found in the lower-resolution FOS spectra. \added[id=AAB]{We incorporate the differing detection limits of the COS and FOS gratings into our analysis in Section \ref{sec:coherence}, and check for any resulting redshift bias throughout this work.}

The catalogue contains 430 absorption systems, of which 272 are \ion{H}{i}. Most of our discussion focuses on these \ion{H}{i} absorbers, although metals are briefly discussed in Section \ref{sec:metals}.

\subsection{Galaxies}\label{sec:gal_data}

\begin{figure*}
\includegraphics[width=\textwidth]{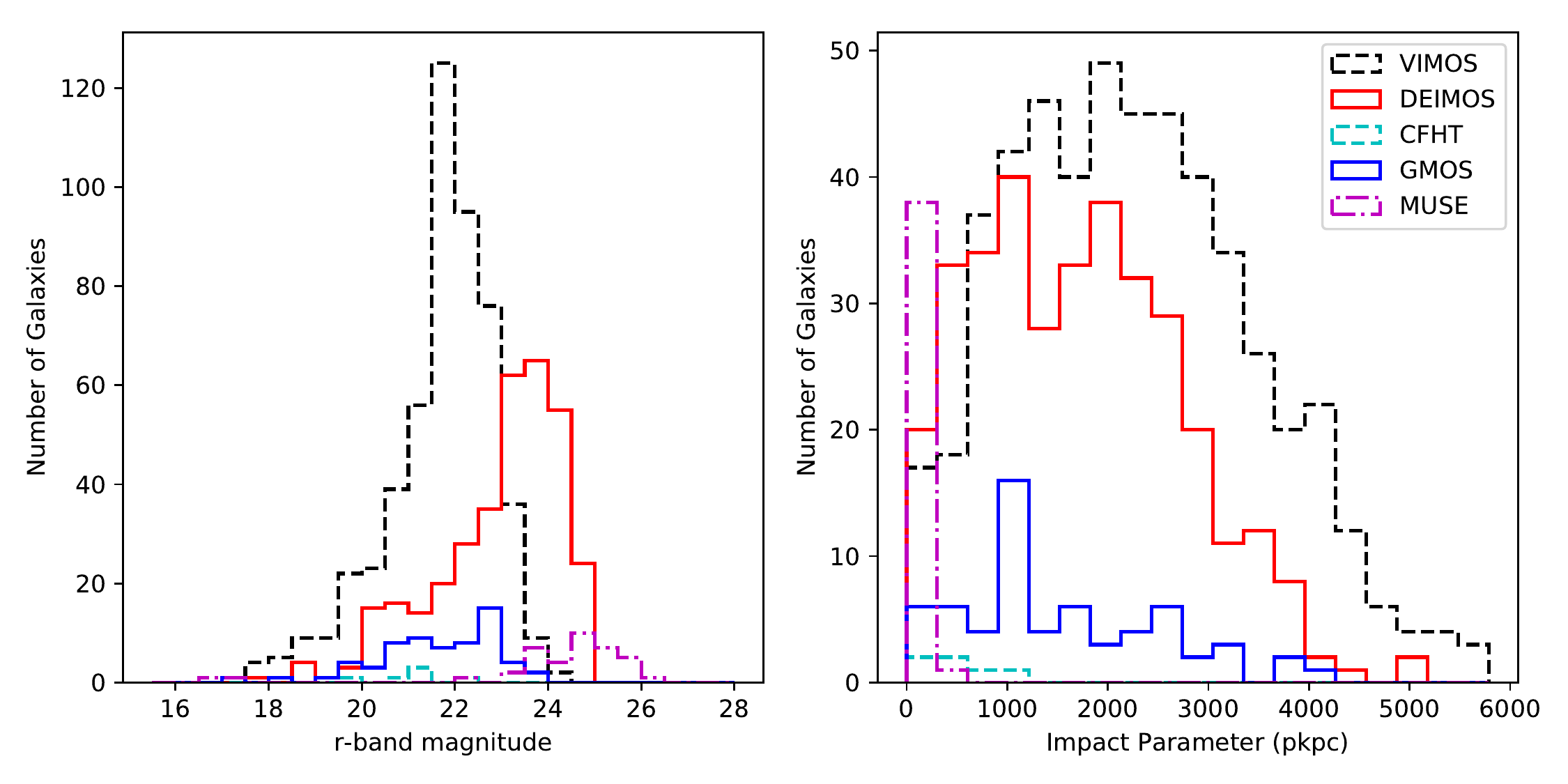}
\caption{The magnitude and impact parameter distributions of galaxies in our catalogue, divided by the instrument used to obtain the galaxy spectrum, illustrating the depth and area of each observing program. Objects observed by multiple MOS instruments are shown only in the instrument with the best redshift flag (these flags are described in Section \ref{sec:z_errors}), or the best resolution if flags are equal. The MUSE objects shown are all new detections not featuring in the MOS surveys. \textit{Left:} The apparent magnitude distribution of the galaxy samples, using the SDSS r-band in bins of 0.5-magnitude width. \textit{Right:} The \added[id=AAB]{impact parameter from each galaxy to} the nearest of the three quasars. \label{fig:mag_hist}}
\end{figure*}

\begin{figure*}
\includegraphics[width=\textwidth]{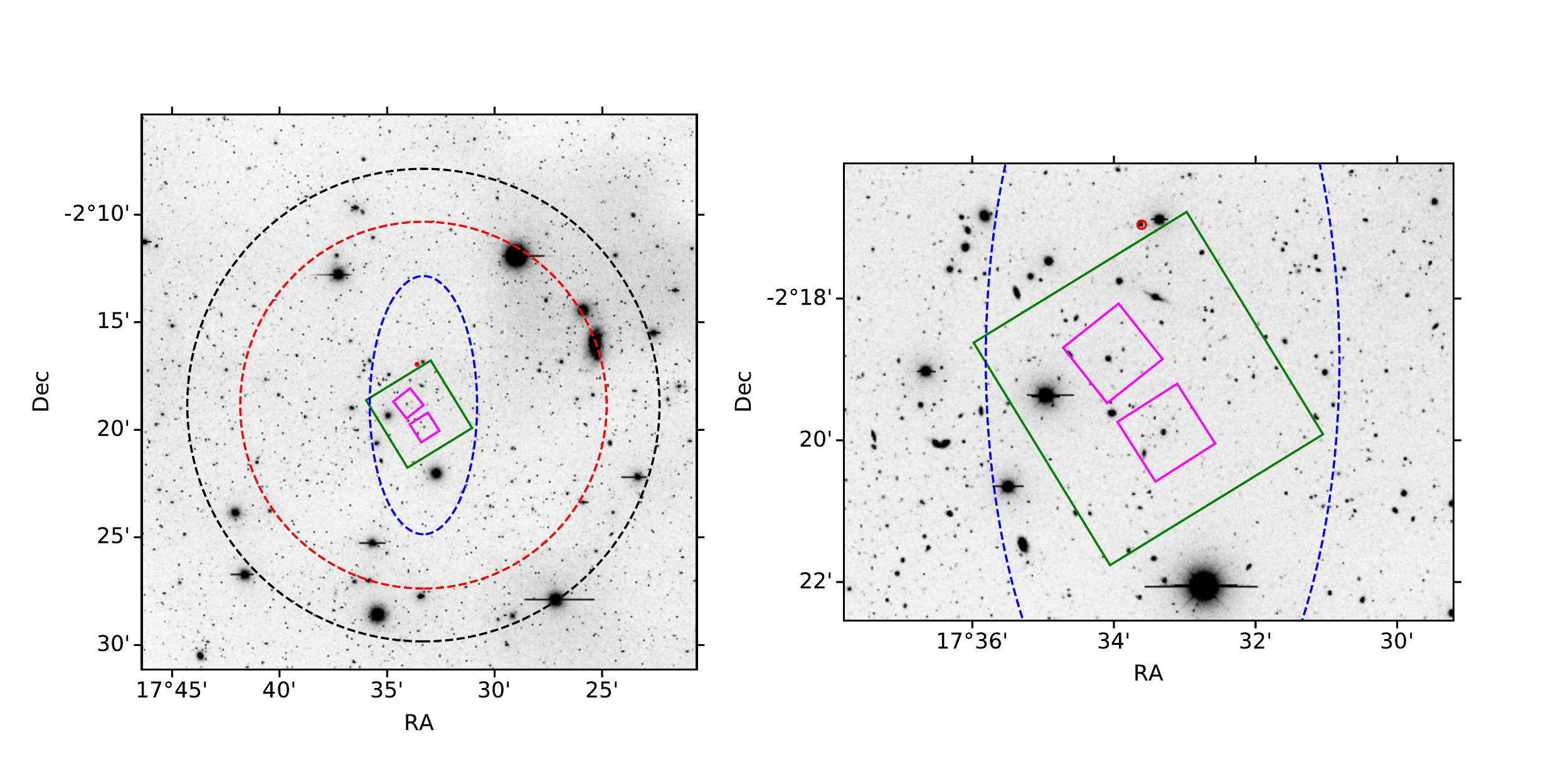}
\caption{The layout of the surveys used in this study. The background image was taken with the Mayall 4-metre Telescope at NSF's Kitt Peak National Observatory. The dashed lines enclose the approximate regions covered by the MOS surveys, with VIMOS, DEIMOS and GMOS shown in black, red and blue respectively. The solid green square shows the region covered by HST imaging, whilst the smaller magenta squares show the MUSE fields centered on QSOs A and B. QSO-C lies inside the small red circle just outside the northern edge of the HST field. The left panel shows the full extent of the galaxy surveys, and the right panel more clearly shows the region close to the lines-of-sight. \label{fig:survey_layout}}
\end{figure*}

The galaxy data used in this study comes from a number of different surveys. The catalogue is based on that used in T14 and F16, with spectra from VIMOS, DEIMOS, GMOS and CFHT-MOS observations (referred to as MOS data throughout). More recent observations using the Multi-Unit Spectroscopic Explorer (MUSE, \hbox{\citealt{bacon2010}}) on the VLT are added to this catalogue. Table \ref{table:mos_summary} summarises the number of spectra taken in each of these observations. Additionally, HST imaging is used to determine position angles and inclinations of galaxies with identified redshifts. These observations and the associated data reduction are discussed below. Figure \ref{fig:mag_hist} shows the magnitude and impact parameter distributions of the MOS and MUSE surveys, whilst their projected extent on the sky is illustrated in Figure \ref{fig:survey_layout}. 


\begin{table*}
\begin{center}
\caption{Summary of the galaxy spectra used in the catalogue. Columns are: (1) instrument used; (2) number of spectra taken (including duplicates); (3) number of unique objects observed by that instrument; (4) number of objects for which that instrument provides the reference spectrum and redshift; (5) number of objects for which we have estimated a redshift using MARZ (used to estimate redshift uncertainties as described in Section \ref{sec:catalogues}); (6) approximate FWHM of the line-spread function in km/s; (7) the program ID of the observations. The instrument with the highest confidence flag is used in the catalogue; if multiple instruments have the same flag, then the best resolution is used.}
\label{table:mos_summary}
\begin{tabular}{| c| r| r| r| r| r| c| }
\hline 
Instrument & N spectra & N unique & N cat & N marz & \added[id=AAB]{FWHM} (km/s) & Prog ID \\
(1) & (2) & (3) & (4) & (5) & (6) & (7) \\\hline

VIMOS & 935 & 757 & 746 & 436 & 1500 & 086.A-0970, 087.A-0857\\
DEIMOS & 642 & 543 & 487 & 286 & 60 & A290D \\
GMOS & 210 & 196 & 112 & 107 & 470 & GS-2008B-Q-50\\
CFHT-MOS & 30 & 29 & 20 & 9 & 400 & \\ \hline
MUSE & 140 & 140 & 59 & 67 & 120 & 094.A-0131\\
\hline

\end{tabular}

\end{center}
\end{table*}


\subsubsection{MUSE} \label{sec:muse}

Information on the kinematics of galaxies close to the quasar lines-of-sight can be extracted from MUSE data. During 2014, the MUSE GTO team took eight exposures covering $1'\times 1'$ fields of view around both QSO-A and QSO-B (program ID 094.A-0131, PI Schaye), totalling two hours for each quasar. This produces a 3D datacube, with a spectrum in each $0.2'' \times 0.2''$ `spaxel' from 4750 to 9350 \AA, with a delivered seeing of $0.96''$ for QSO-A, and $0.82''$ for QSO-B. In the spectral direction, the datacube has a FWHM of $\approx$2.7 \AA.

The reduction of these data follows a similar process to \citet{fumagalli2016, fumagalli2017, fossati2019, lofthouse2020, bielby2020}. MUSE ESO pipeline routines were used to remove bias, apply flat-fielding, and calibrate astrometry and wavelength for each exposure. The `scibasic' and `scipost' pipeline routines combine the IFUs for each exposure, resampling using a drizzle algorithm onto a 3D grid, as well as correcting for telluric absorption using sky continuum and sky line models produced using the darkest pixels in the exposure. Exposures can be aligned using point sources to produce a reasonable combined datacube, but this process generally leaves sky line residuals as well as uneven illumination across the field \citep[e.g.][]{bacon2017}.

Further corrections were applied using the CUBEX package (S. Cantalupo, in prep). We use two main routines from this package \citep{cantalupo2019}. CubeFix performs a renormalization on each IFU, stack, and `slice' (similar to a single slit), making the background as flat as possible across the MUSE field-of-view and wavelength range, and removing the `chequered' pattern that often afflicts images reduced solely using the ESO pipeline. CubeSharp provides a flux-conserving sky subtraction, using the empirical shape of sky lines to calculate a line-spread function. \added[id=AAB]{Within each spatial pixel, flux is then allowed to move} between neighbouring spectral pixels to best match the LSF, allowing the sky lines to be more accurately removed \added[id=AAB]{whilst conserving total flux}. CubeFix and CubeSharp are run twice on each exposure, the results of the first run allowing better masking of sources in the second run. This further ensures that fluxes are not overcorrected and preserves the source flux as well as possible.

A 3$\sigma$ clipping is then used, combining the exposures using mean statistics. This combined cube is then used to mask sources for a final run of CubeFix and CubeSharp. Using this package greatly reduces sky and flat-field residuals, although some residuals remain visible towards the red end of the MUSE spectra.

Objects in the MUSE fields were identified using SExtractor \citep{bertin1996} on the white-light image. We produced 1D spectra by summing the flux within the SExtractor aperture.

We then estimated redshifts using the MARZ software \citep{hinton2016}, with additional galaxy templates provided by Matteo Fossati (described in \citealt{fossati2019}, created using \citealt{bruzual2003} stellar population models).  MARZ identifies the five redshift/template combinations producing the best cross-correlation between the observed spectrum and the template. We then chose the most likely of these based on the features fitted. Objects identified by SExtractor that are much smaller than the point-spread function ($\approx$1", 5 pixels) can be excluded as artifacts, and other objects are best fit by a stellar template instead of a galaxy template. The remaining objects have been assigned a confidence flag between 1 (redshift unknown) and 4 (highly confident) based on the spectral features visible at the redshift given by the best-fit result from MARZ. 67 galaxies were assigned a flag $> 1$, and are therefore used in this study. The number of objects with each confidence flag is shown in Table \ref{table:muse_flags}.

\begin{table}
\begin{center}
\caption{Objects detected in the MUSE fields by the MARZ flag assigned. The flag 2-4 galaxy detections are those added to the galaxy sample used for this study.}
\label{table:muse_flags}
\begin{tabular}{| c| c| r|}
\hline 
Flag & Descriptor & N \\ \hline

4 & highly confident & 24  \\
3 & good & 21  \\
2 & possible & 22 \\  \hline
1 & unknown & 39\\
6 & artefacts/stars & 32  \\
 & QSOs & 2 \\ \hline
2-4 & galaxy detections & 67 \\
 & total & 140 \\\hline

\end{tabular}

\end{center}
\end{table}

\subsubsection{MOS} \label{sec:mos}

\begin{figure*}
\includegraphics[width=\textwidth]{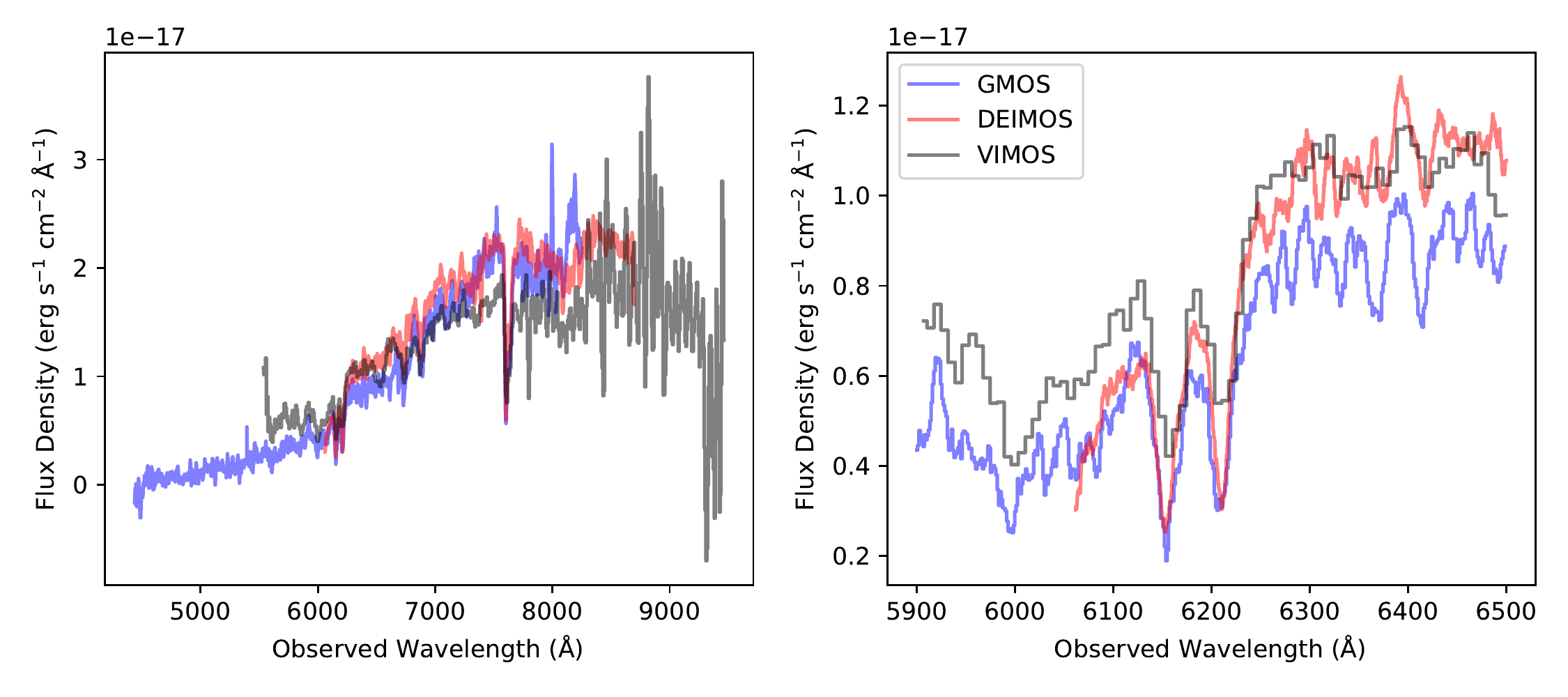}
\caption{Example spectra of a galaxy observed in VIMOS, GMOS and DEIMOS. The left panel shows the full spectrum highlighting the different wavelength ranges covered, whilst the right panel centers on the Ca K \& H absorption features. This galaxy is at z $\approx$ 0.56, has luminosity close to $L_{\star}$, and is classified as non-star-forming. We smooth over a 3 \AA{ }kernel to improve visibility. \label{fig:ex_spec}}
\end{figure*}

The MOS galaxy data is the subset of the catalogue from T14 that covers the Q0107 field, consisting of spectra from CFHT-MOS, VIMOS, DEIMOS and GMOS. Many objects in the catalogue were observed multiple times, either by the same instrument or by multiple instruments. One example is shown in Figure \ref{fig:ex_spec}. The data are described briefly here. T14 and references therein describe the data collection and reduction processes in more detail. 

\subsubsection{CFHT-MOS}

The multi-object spectrograph on the Canada-France-Hawaii Telescope (CFHT-MOS) \hbox{\citep{lefevre1994}} was used for observing runs in 1995 and 1997 by \hbox{\citet{morris2006}}. The Q0107 field was observed on the 29th and 30th July 1995, producing 30 galaxy spectra (one object observed twice, so 29 objects). The observation and reduction are described in more detail in the above paper. 

The observed spectra were bias-subtracted using IRAF, and bad pixels were interpolated over. Cosmic rays were removed by comparing multiple exposures using the same mask. Sky subtraction used adjacent regions of the slit, whilst wavelength calibration used an arc frame obtained whilst pointing towards the same region of sky (to minimize the effects of instrument flexure). Flux calibrations used a nearby standard star. 

The number of CFHT galaxies observed is not sufficient to find a statistically significant offset between these and other observations, and no redshift confidence flags were provided.

\subsubsection{VIMOS}

The VIMOS data \citep{lefevre2003} used a low-resolution (R $\approx$ 200) grism, giving 935 spectra with coverage between 5500 and 9500 \AA{ }(programs 086.A-0970, PI:Crighton; and 087.A-0857, PI: Tejos). The data were reduced using VIPGI \citep{scodeggio2005}. Wavelength corrections were made using both lamp frames and skylines, whilst flux calibration used a standard star. \added[id=AAB]{We note that these data were taken shortly before the VIMOS charge-coupled devices (CCDs) were updated in August 2010, and are unfortunately affected by fringing effects at wavelengths $\gtrsim$7500 \AA.}

Redshifts were estimated using cross-correlation between the observed spectra and SDSS templates. These redshifts were then manually assigned a confidence flag (a: secure, b: possible, c: uncertain), where secure redshifts required at least three spectral features. The redshifts of all MOS objects in the Q0107 field were then adjusted to match the DEIMOS frame (as DEIMOS has the best resolution of the MOS instruments used), based on the objects observed by VIMOS and DEIMOS. The magnitude of this shift was $\Delta z \approx$ 0.0008, or 120-240 $\textrm{km}$ $\textrm{s}^{-1}$ (see T14 for details).

\subsubsection{DEIMOS}

The DEIMOS \citep{faber2003} settings give a much better resolution (R $\approx$ 5000) and substantially deeper data but over a smaller field, covering the 6400-9100 \AA{ }range for 642 objects (taken in 2007-08, program A290D, PIs: Bechtold and Jannuzi). Redshifts were obtained from the DEEP2 data reduction pipeline \citep{newman2013}, which applied all necessary de-biasing, flat-fielding, wavelength and flux calibration, and heliocentric corrections. The DEIMOS redshift confidence was measured in the pipeline using four categories (1: not good enough, 2: possible, 3: good, 4: excellent), which were reassigned to match the three categories above (1 to c, 2 and 3 to b, 4 to a) when added to the catalogue.

\subsubsection{GMOS}

GMOS \citep{davies1997} was used in 2008 on this field (program GS-2008B-Q-50, PI: Crighton), with an intermediate resolving power (R $\approx$ 640) and a slightly bluer wavelength range of 4450--8250 \AA. 

Each spectrum consists of three 1080s exposures, dithered in wavelength to allow removal of bad pixels. IRAF was used to calibrate fluxes and wavelengths, using arc frames and a standard star taken contemporaneously with the science exposures. 

210 redshifts were estimated using the same method as for VIMOS data, although the shift needed to match the DEIMOS frame was smaller, only $\Delta z \approx$ 0.0004 or 60-120 $\textrm{km}$ $\textrm{s}^{-1}$.

\subsection{Combined galaxy catalogue}\label{sec:catalogues}

In order to combine spectra from the multiple instruments previously described into a single galaxy catalogue for this field, we need to match objects observed by multiple instruments in order to remove duplicates, as well as ensure that the galaxy properties we utilise in our analysis are measured consistently. We use photometry and astrometry from the Sloan Digital Sky Survey source catalogue (SDSS, \citealt{albareti2017}) as an anchor, as the SDSS catalogue includes close to half of the MOS objects, and 14 of the MUSE galaxies. 

\subsubsection{Astrometry}\label{sec:astrometry}

In order to remove duplicates and select a single spectrum from which to derive the properties of each galaxy, we matched the coordinates of objects observed by multiple MOS instruments to those observed in SDSS. With the exception of CFHT, all instruments had sufficient cross-matches within 1$"$ to confirm that the astrometry is consistent between the MOS instruments, and required an offset of less than $0.5"$ to match SDSS. The same process applied to the MUSE astrometry revealed a similar $\approx 1"$ offset for both fields, which were corrected separately. 

The astrometry of both the MOS and MUSE catalogues were adjusted to match SDSS, to ensure objects appearing in both catalogues were correctly paired. 28 objects were found to match within 1" after the correction to SDSS, and no additional matches within 2". Due to the larger number of objects, the offset between MOS and SDSS is more accurately determined than that for MUSE, so the corrected MOS coordinates are used for the objects appearing in both catalogues.

\subsubsection{Magnitudes}\label{sec:magnitudes}

R- and I- band magnitudes included in the MOS catalogue are systematically shifted to match the r- and i- band SDSS magnitudes, using the Lupton (2005) transformations\footnote{ \url{https://www.sdss.org/dr16/algorithms/sdssubvritransform}}. Magnitudes for MUSE-only objects were estimated by integrating the spectrum through the SDSS r- and i- filters. The corrected MOS magnitudes were preferred for objects appearing in both MOS and MUSE catalogues, and the differences were approximately consistent with the uncertainties provided.

\added[id=AAB]{The survey depths are limited by the original target selections, at R $\approx$ 23.5, 24 and 24.5 for VIMOS, GMOS and DEIMOS respectively. As shown in Figure 8 of T14, the sucess rate of assigning redshifts to objects in the Q0107 field is $\gtrsim$ 80\% to a depth of R=22 in both VIMOS and DEIMOS, with the deeper DEIMOS data showing $\gtrsim$ 60\% success to R=24.}

\added[id=AAB]{We do not detect any significant variation in the depth of the MOS data within the HST field, although the small number of MUSE galaxies extend to fainter objects. Any variation in depth across the field does not have a significant effect on our results, as we retain the same selection for our analysis in Section \ref{sec:coherence}, and our Section \ref{sec:orientations} and \ref{sec:kinematics} use the much smaller HST field.}

\subsubsection{Redshifts}\label{sec:redshifts}

Calculations of relative velocities between galaxies and absorbers require that there are no systematic shifts between their redshifts. We therefore compared redshifts to ensure that all of our galaxy samples are in the same frame as the absorption features. 

The separate MOS samples from T14 have already been corrected to a single frame, described in the section for each instrument. We therefore first check that this galaxy frame is the same as the frame in which the absorber redshifts are measured. One test of this is to calculate the difference in redshift between galaxy--absorber pairs. At the scales on which galaxies and absorbers are correlated, this should produce a signal that is symmetric about zero (velocity offsets e.g. outflows should average to zero over a large sample), as presented by \citet{rakic2011}. If the signal is offset, that suggests a shift is needed to bring all the redshifts into the same frame. 

We find no such shift when pairing high column density \ion{H}{i} absorbers (N(\ion{H}{i}) $>$ $10^{14}$ $\textrm{cm}^{-2}$) with galaxies within 2.5 Mpc of at least one sightline, both from the T14 catalogues. We fit a model consisting of a Gaussian plus constant offset to the
velocity distribution of these pairs, these two components representing associated absorbers with small velocity differences, and unconnected absorbers with uniformly distributed velocity differences. 

\added[id=AAB]{As most possible pairs lie in the uniform distribution, a velocity cut is needed to centre the fit close to zero, and avoid noise in this distribution dominating over the Gaussian peak (i.e the Gaussian component of the fit jumps to a nearby `noise spike' rather than the peak of associated galaxy--absorber pairs).} The fit varies with
\added[id=AAB]{the velocity used for this cut}, but the centroid of the Gaussian component remains within $\pm 10$ $\textrm{km}$ $\textrm{s}^{-1}$ for 
\added[id=AAB]{cuts smaller than 5000 $\textrm{km}$ $\textrm{s}^{-1}$}. Therefore the galaxies and absorbers provided by T14 can be taken to lie in the same frame. We similarly confirm that the separate samples for each instrument in the T14 catalogue are each consistent with this frame.

A small offset can be found if weaker absorbers are used or galaxies with a larger impact parameter are included \added[id=AAB]{, as well as if a larger velocity cut is used}. These are less likely to be physically connected, so introduce noise that dominates over the peak
We therefore do not 
\added[id=AAB]{attempt to correct for any such} spurious offset.

We then add the MUSE observations. The distribution of velocity differences between MOS and MUSE observations of duplicate objects showed that the MUSE objects required a further shift of 30 $\textrm{km}$ $\textrm{s}^{-1}$ in order to match the frame of the MOS catalogue.

\begin{figure*}
\includegraphics[width=\textwidth]{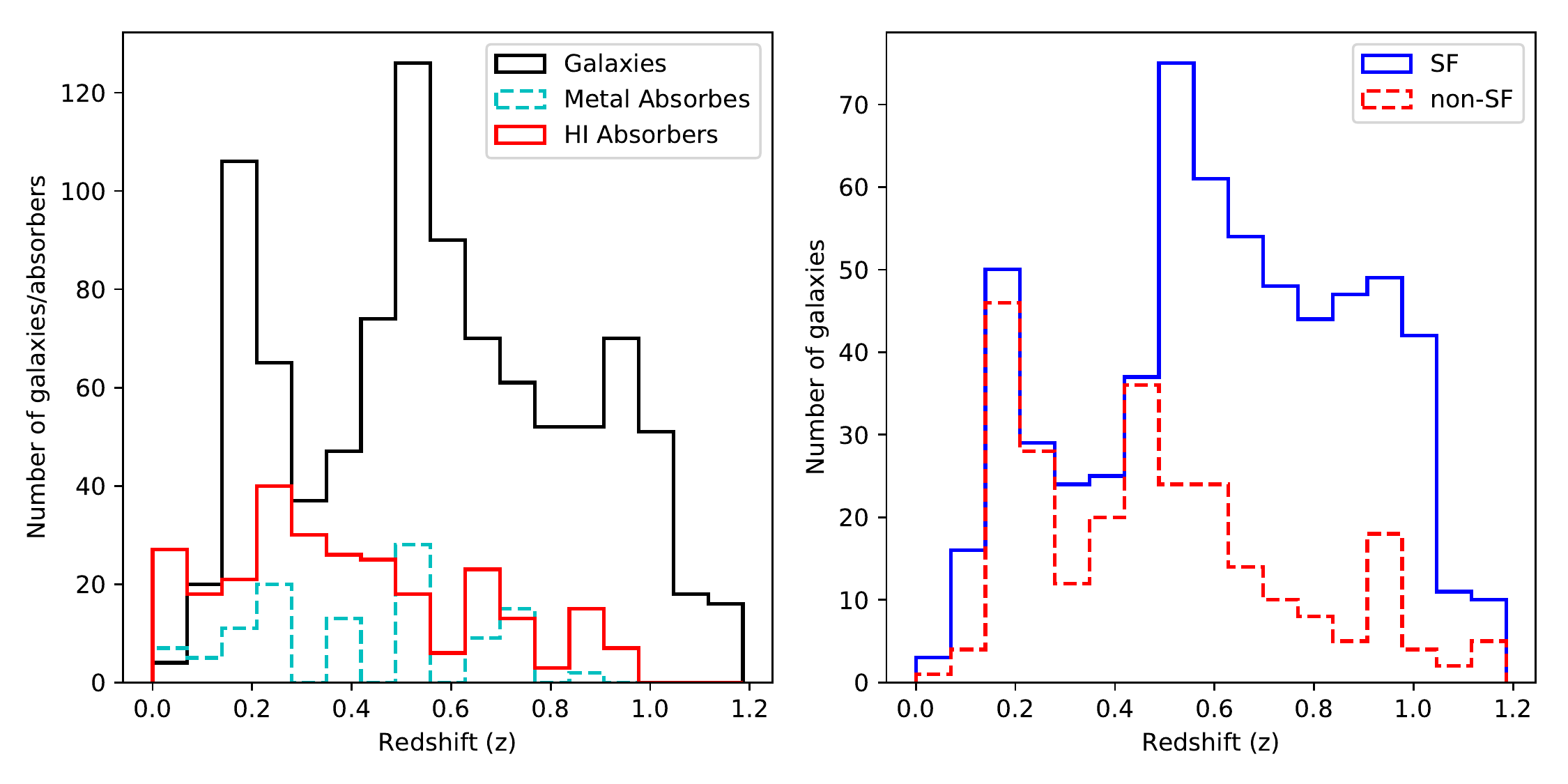}
\caption{\added[id=AAB]{The redshift distributions of galaxies and absorbers in our catalogue. \textit{Left:} The redshift distributions of the galaxy and absorber samples. \textit{Right:} The redshift distributions of the star-forming and non-star-forming galaxies in our sample.} \label{fig:z_hists}}
\end{figure*}

\added[id=AAB]{We note that, although the redshifts of galaxies and absorbers are in the same frame, the overall redshift distributions are not similar, as seen in the left panel of Figure \protect\ref{fig:z_hists}. Therefore, when comparing the galaxy--absorber associations involving different sub-samples, the difference in redshift must be considered in our analysis.}

\subsubsection{Redshift uncertainties}\label{sec:z_errors}

Redshift uncertainties for the MOS objects are based on those with multiple spectra. We attempted to fit all MOS spectra using the MARZ code used earlier to obtain redshifts from MUSE. For objects with multiple spectra from the same instrument, the redshift differences could be compared. The width of the distribution of velocity differences for each instrument provides an estimate of the uncertainty in the redshift of objects observed by that instrument. These are given in Table \ref{table:z_errors}, where $\sigma$ is adopted as the uncertainty for all objects in the T14 catalogue with the confidence flag shown in brackets. As redshifts with poorer confidence flags are generally identified using fewer features, we assign a larger redshift uncertainty to these b-flag galaxies. Whilst the GMOS sample size is very small, the relative values of the three instruments are consistent with the resolutions given in Table \ref{table:mos_summary}, so we adopt these uncertainties. 

Our MUSE observations contain no duplicates, so this method cannot be used to estimate uncertainties on their redshifts. As MUSE has a higher resolution than GMOS, but lower than DEIMOS, we take the GMOS values as estimates of the velocity errors for MUSE galaxies with the same confidence flags.

\begin{table}
\begin{center}
\caption{Comparison of the measured velocity differences between objects measured twice by the same MOS instrument. We show the lowest-confidence flag assigned to spectra of each object from each instrument, and consider the distribution of velocity differences of each sample. Redshifts and flags shown are those assigned using the MARZ software (flags from 1 being `unknown' to 4 being `highly confident'), and we also show the flags given by the T14 catalogue to which the MARZ flags are matched. A velocity cut of 2000 $\textrm{km}$ $\textrm{s}^{-1}$ was used to remove pairs of observations for which the velocity difference is due to a different identification of spectral features. The cut affects only a small number of `flag 2' objects. The number removed by this cut is shown in brackets. $\sigma$ is the standard deviation of the velocity differences, which can be adopted as the uncertainty on any individual redshift measurement within that sample.}
\label{table:z_errors}
\begin{tabular}{| c| c| c| r| }
\hline 
Intstruments & Flags & N & $\sigma$ (km/s) \\ \hline 

VIMOS & 3, 4 (a) & 29 & 120 \\
 & 2 (b) & 69 (6) & 190 \\ \hline
DEIMOS & 3, 4 (a) & 23 & 26 \\
 & 2 (b) & 40 (3) & 47 \\ \hline
GMOS & 3, 4 (a) & 4 & 44 \\
 & 2 (b) & 5 & 48 \\ \hline

\end{tabular}

\end{center}
\end{table}

In order to directly compare the flags assigned for objects with MOS and MARZ redshifts (a--c in the catalogue, 4--1 in MARZ), flag `a' objects were numbered 3.5, `b' to 2.5 and `c' to 1.5. Redshifts quoted in our final catalogue are those with the highest flag.

Finally, the MARZ flags were reassigned to match those from the MOS catalogue: 1 to c, 2 to b and 3 \& 4 to a. This produces our final catalogue of 1424 galaxies, of which 1026 have `a' or `b' flags and are used in the following analyses. Their locations in space and redshift are shown in Figure \ref{fig:wedge_plot}.

\subsubsection{Spectral Classification} \label{sec:sf_classes}

We divide our galaxy sample into `star-forming' and `non-star-forming' galaxies. We maintain the classifications used in T14 for the MOS galaxies (see their section 5.1) and apply similar criteria for dividing the MUSE galaxies. Namely, those galaxies best fit by a star-forming template (e.g. `late-type', `starburst' or `star-forming' templates) are classified as star-forming, and those fit by a passive template (e.g. `passive', `early-type' or `absorption galaxy' templates) are classified as non-star-forming. 

These templates differ primarily due to the presence of emission lines, so this classification is denoting galaxies with measurable emission lines as star-forming, and those without as non-star-forming. 

\begin{figure}
\includegraphics[width=\columnwidth]{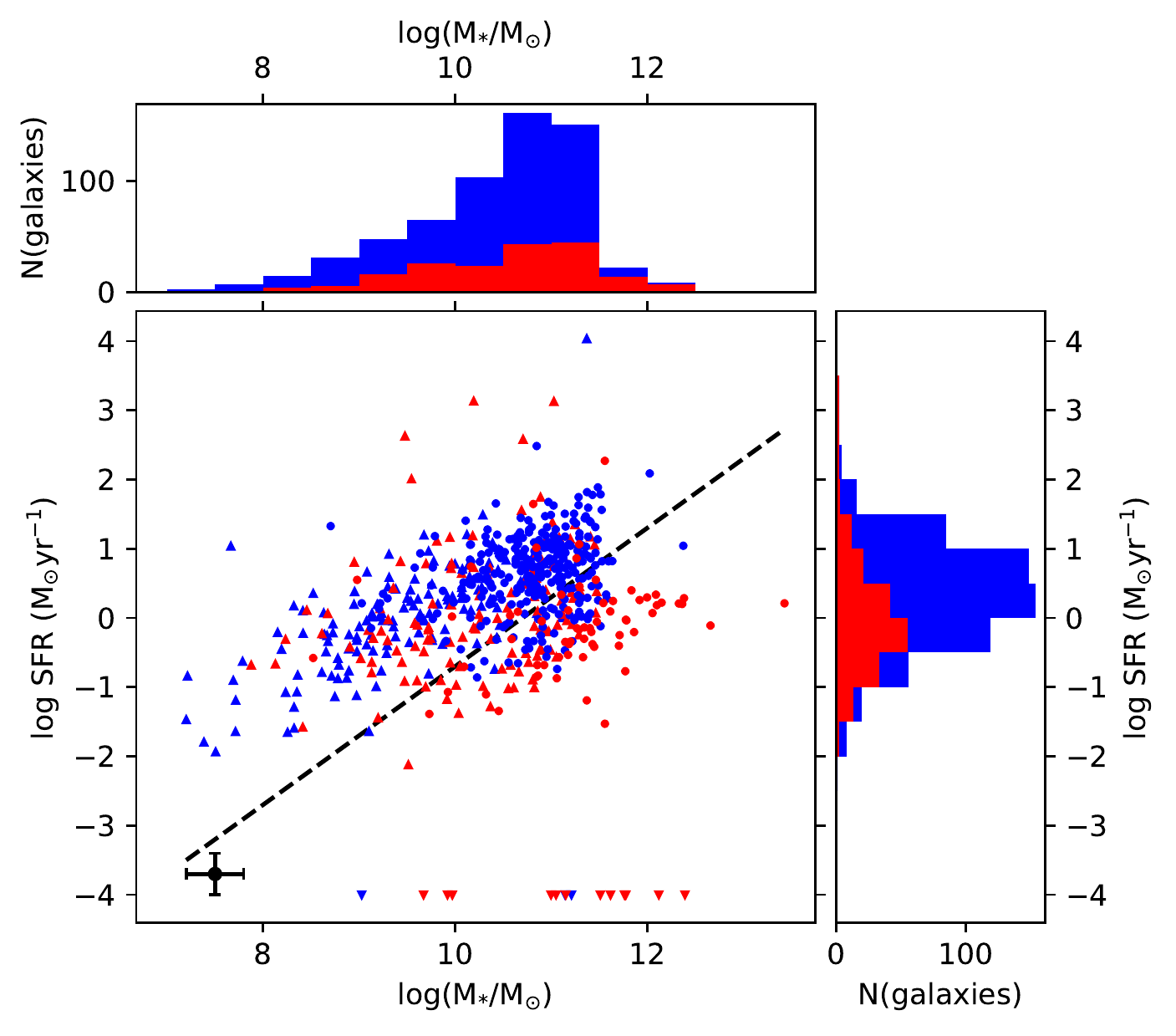}
\caption{\added[id=AAB]{The star-formation rate against stellar mass for our galaxy sample. Star-formation rates are calculated using the [OII] or H$\alpha$ emission lines, shown in circles and upward-pointing triangles respectively. The downward-pointing triangles at the bottom of the figure show the masses of galaxies for which the line fit was unable to produce a result. (These are excluded from both projected histograms.) The blue points and histograms denote galaxies classified as `star-forming' using the template fits, whilst the red points were classified as `non-star-forming'. The black point in the lower-left shows the median uncertainty in each axis, although the non-star-forming galaxies exhibit systematically larger uncertainty in SFR. The dotted black line is the 0.02 $\textrm{Gyr}^{-1}$ cut in sSFR that best reproduces our template classification. Most of the objects lying substantially above the apparent `main sequence' are due to the emission line being masked by sky lines or the fringes seen in the VIMOS spectra at red wavelengths. \protect\label{fig:SFR_mass}}}
\end{figure}

\begin{figure*}
\includegraphics[width=0.79\textwidth]{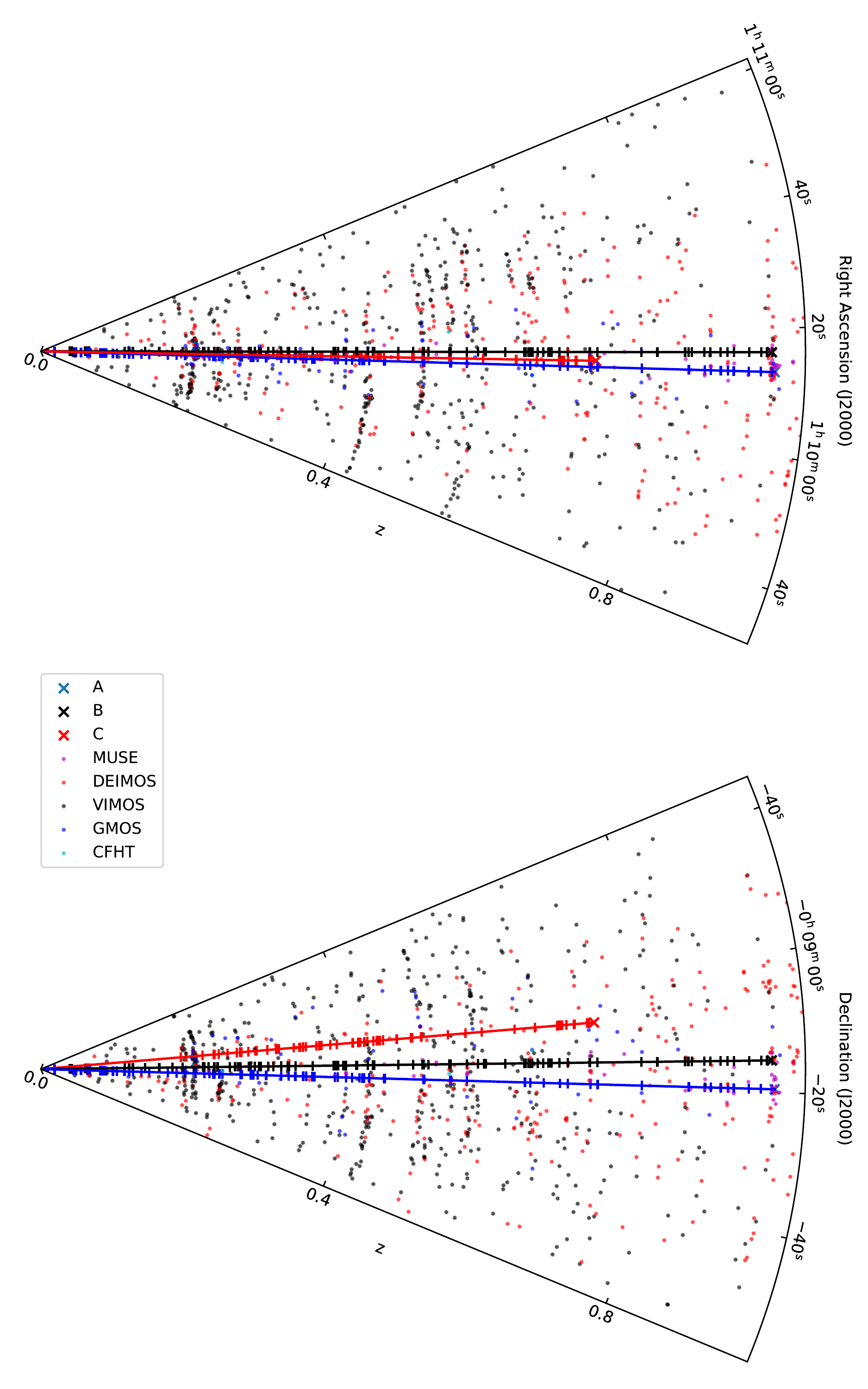}
\caption{The distribution of galaxies and \ion{H}{i} absorption features in the sample, plotted in redshift and right-ascension (top) or declination (bottom). Solid lines show the lines-of-sight to the three quasars, with HI absorption shown as tick marks. Points are galaxies coloured by observed instrument. Note that the angles are massively expanded, so arc-like features likely correspond to galaxy groups. \label{fig:wedge_plot}}
\end{figure*}

We also estimate the star-formation rates by directly fitting the H$\alpha$ and [\ion{O}{ii}] emission lines\footnote{We use the \citet{kennicutt1998} \added[id=AAB]{and \citet{kewley2004} calibrations to convert line luminosity to SFR for H$\alpha$ and [\ion{O}{ii}] respectively.} We assume 1 magnitude of extinction at H$\alpha$ \citep{charlot2002}, \added[id=AAB]{and use the \citet{calzetti2000} curve to estimate extinction at [OII]}. The predicted wavelength of at least one of these emission lines is available for $\approx$ 90\% of galaxies with well determined redshifts.}, and estimate stellar masses using the \added[id=AAB]{relationship given by \citet{johnson2015}}
, finding that a cut at a specific star-formation rate of 0.02 $\textrm{Gyr}^{-1}$ correctly identifies $\approx$75\% of both samples. \added[id=AAB]{Our stellar mass and SFR estimates are illustrated in Figure \protect\ref{fig:SFR_mass}}. However, these estimates have substantial measurement and systematic uncertainties, so we use the binary classification in our analysis.

These subsamples of star-forming (SF) and non-star-forming (non-SF) galaxies show no substantial bias in their impact parameter distributions or mass distributions. It must be noted that there is a small excess of star-forming galaxies at the smallest impact parameters ($< 200$ kpc), due to the ease of finding emission lines using the MUSE datacubes. The SF and non-SF samples also feature slightly larger low-mass and high-mass tails respectively in their mass distributions, but both of these biases affect a small number of galaxies. We confirm throughout that these have no substantial effect on our results by re-running our tests with samples excluding the MUSE galaxies and samples excluding the tails of the mass distribution, obtaining similar results.

However, there is a substantial bias in redshift, with SF galaxies preferentially featuring at higher redshifts than non-SF galaxies \added[id=AAB]{, as seen in the right panel of Figure \ref{fig:z_hists}}. This is likely a combination of real redshift evolution (higher cosmic star-formation rates at higher redshift, e.g. \citealt{madau2014}) and observational effects (higher signal-to-noise is required to confirm the redshifts of galaxies without emission lines), and must be taken into account when comparing the CGM/IGM around galaxies in these samples.

\subsection{HST imaging} \label{sec:HST}

In addition to the spectroscopic data, we also use high-resolution imaging of the field to constrain galaxy morphologies and orientations. We use publicly available Hubble Space Telescope imaging of this field (Program ID: 14660, PI Straka), obtained through ACS \citep{ryon2019a} and the F814W filter. This consists of four exposures totalling 2171 seconds. 

It must be noted that one of the exposures was affected by an unidentified bright object moving across the field, leaving an artefact in the final combined image. However, we only use the HST imaging to study the morphology of galaxies in this field, so this artifact does not substantially affect our results.

Galaxies were identified using SExtractor, then matched to the coordinates of objects in the MOS/MUSE combined catalogue. No systematic offset was found, so objects within 1$"$ were matched, as above.

We run GALFIT \citep{peng2002}, which uses chi-squared minimization algorithms to produce a best-fitting 2D model of a galaxy. We initially fit a Sersic disk to every galaxy found in both our redshift catalogue and the HST image, using SExtractor results as initial guesses for the fit, and then introduce additional components where necessary to find a reasonable fit. This provides improved position angles and inclinations, taking full account of the point-spread function of the image and reducing the average uncertainty by a factor $\approx$ 3 relative to the position angles produced by SExtractor. 

We again assign quality flags to the GALFIT results \added[id=AAB]{(1: good fit by eye and no clear structure remains in residuals, 2: good by eye, 3: possible, 4: clearly a poor fit )}, allowing poorly constrained results to be excluded. Flag 4 objects are excluded from all results. This returns 109 galaxies with 
position angles \added[id=AAB]{constrained by GALFIT}
\added[id=AAB]{and a counterpart in our spectroscopic survey, of which 72 also have well-constrained redshifts (`a' or `b' flags). We illustrate examples of this fitting in Appendix \ref{sec:galfit_results}.}

\section{Coherence Between sightlines} \label{sec:coherence}

One unique test allowed by the configuration of multiple lines-of-sight is to compare the absorption seen across multiple sightlines at the same redshift. Both \citet{dinshaw1997} and \citet{crighton2010} (C10) attempted to estimate the scale size of absorbers using the numbers of coincidences (where absorption was identified at multiple redshifts) and anti-coincidences (where detection of absorption in one sightline was not matched in the other(s)). We can therefore use our larger sample to both review the results from these papers, demonstrating that absorption is often correlated on the 400-1200 kpc scales separating these sightlines, and to split the sample, allowing us to study how these coincidences are affected by the properties of the absorption and of nearby galaxies.

\subsection{Random absorbers}

\begin{figure}
\includegraphics[width=\columnwidth]{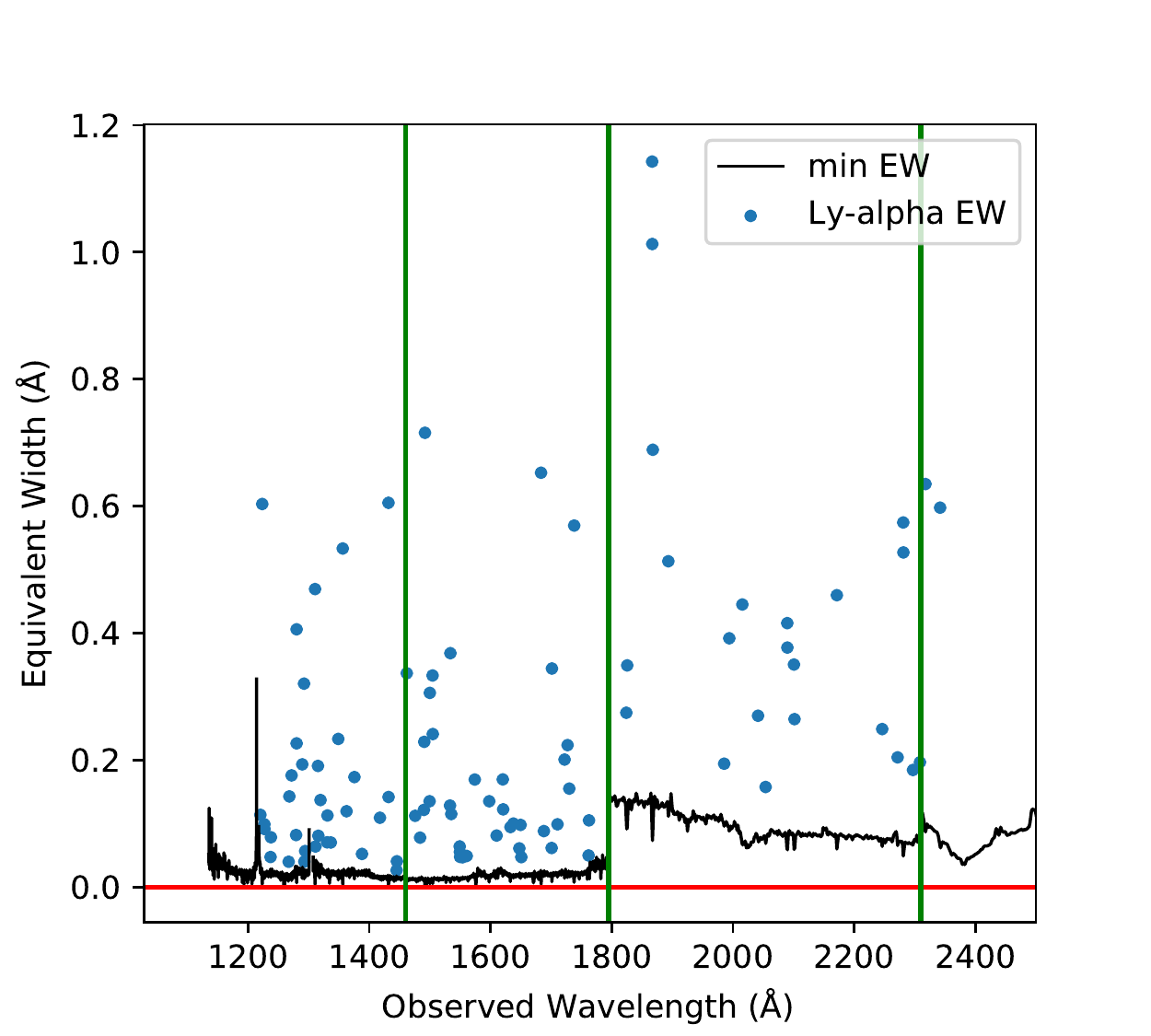}
\caption{The estimated minimum rest-frame equivalent width detectable in the COS and FOS spectrum of QSO-A. The calculation is described in the text, SNR is calculated using the continuum estimate and the instrumental noise, with the FWHM taken from Table 2 in T14 for each grating. A significance limit of 3 is used. The observed \ion{H}{i} absorbers from the catalogue are scattered on the plot, assuming that the Ly-$\alpha$ line was observed. The vertical green lines divide the different gratings used in the spectrum, which therefore have different signal-to-noise and FWHM.\label{fig:ew_detection}}
\end{figure}

In order to test the significance of the correlations between sightlines, we must compare the observed distribution to the number expected if there were no physical connection between the observed gas and galaxies. For this reason, we have generated 5000 sets of randomly distributed absorbers, using a method similar to that used in T14, as follows:

\begin{enumerate}
 \item Calculate the signal-to-noise per resolution element for each QSO spectrum.
 \item Convert this to a minimum rest-frame equivalent width for the absorption feature as a function of redshift. The detection limits for QSO-A are shown as an example in Figure \ref{fig:ew_detection}.
 \item For each real absorption feature, find the allowed region in redshift space for which the EW of the progenitor is larger than the minimum, and is not covered by galactic absorption.
\item Distribute absorbers randomly through the allowed region,  giving the random absorber the same properties as the observed progenitor.
\end{enumerate}

\added[id=AAB]{In order to maintain the approximate redshift/wavelength distribution of absorbers, we restrict random absorbers to the same grating as their observed progenitor.}
We describe this process in more detail in Appendix \ref{sec:random_absorbers}.

\subsection{Absorber-Galaxy Groups}

In order to associate galaxies and absorbers, we use two different grouping algorithms in this section, similar to those used in C10. The first is a nearest-neighbour algorithm, which simply takes the nearest absorber (in velocity) to each galaxy in each sightline. 

The second method is a velocity-cut around each absorber/galaxy, in which we consider all absorbers/galaxies within that velocity window as associated with that absorber/galaxy. 

Galaxies must be within 2.5 Mpc of at least one of the sightlines in order to be included, and have a redshift flag of `a' or `b'. These restrictions remove the galaxy-absorber pairs with the largest separations (and are therefore the least likely to be physically connected), as well as those with poorly-determined redshifts. 

\subsection{Results} \label{sec:agplots}

In the series of Figures \ref{fig:randoms_base_v500}, \ref{fig:randoms_coldens} and \ref{fig:randoms_starform}, three plots are shown for each set of constraints applied to the samples, each with three panels. The top three panels show the velocity difference between each galaxy and its nearest-neighbour absorber in each sightline. The histogram shows the distribution of real velocity differences, with the black points giving the median number of galaxies in each bin in the random sets, and the cyan points showing the 99\% level. Thus, the level of the histogram in relation to the points shows the excess of galaxy-absorber separations in that 100 $\textrm{km}$ $\textrm{s}^{-1}$ bin over the expectation if the absorber redshifts were randomly distributed. Note that, as the total number of galaxies is the same, any excess in bins with small velocity difference must be accompanied by a deficit in other bins.

The middle three panels show the number of galaxies around which at least 1, at least 2, or all 3 sightlines contain \ion{H}{i} absorption within the velocity-cut. The histogram shows the distribution of the random sets, with the black vertical line giving the mean value. The red vertical line shows the number of galaxies found in the real system. Also given are the percentage of random sets in which the number of galaxies found with absorption in 1, 2 or 3 sightlines is greater than or equal to the number in the real Q0107 system, and the significance of the difference from the mean in units of the standard deviation of the random distribution. Therefore panels in which the red line lies to the right of the histogram show that there are more galaxy--absorber groups in the real Q0107 system than in the random distributions.

The final panels show the number of \ion{H}{i} absorber groups for which absorption within the velocity window is found in precisely 1, 2, and 3 sightlines. The layout of the plot is as described above, with the percentage of random sets containing at least as many absorber groups as the real system given alongside the significance of the difference. The uppermost of these three panels often shows the real system having fewer single-LOS absorber groups than the random distribution, as the real absorbers are more likely to form coherent structures across multiple lines-of-sight, and be more clustered along any one line-of-sight.

These figures are intended as a direct comparison with those in C10. However, with the much larger samples now available, it is possible to obtain results from subsamples of galaxies and absorbers. These include separating star-forming from non-star-forming galaxies, and dividing absorption systems into low- and high-column-density samples. This allows testing of numerous models of the links between the galaxies and surrounding gas, as described below.

\subsubsection{Full Sample} \label{sec:base}

\begin{figure}

\includegraphics[width=\columnwidth]{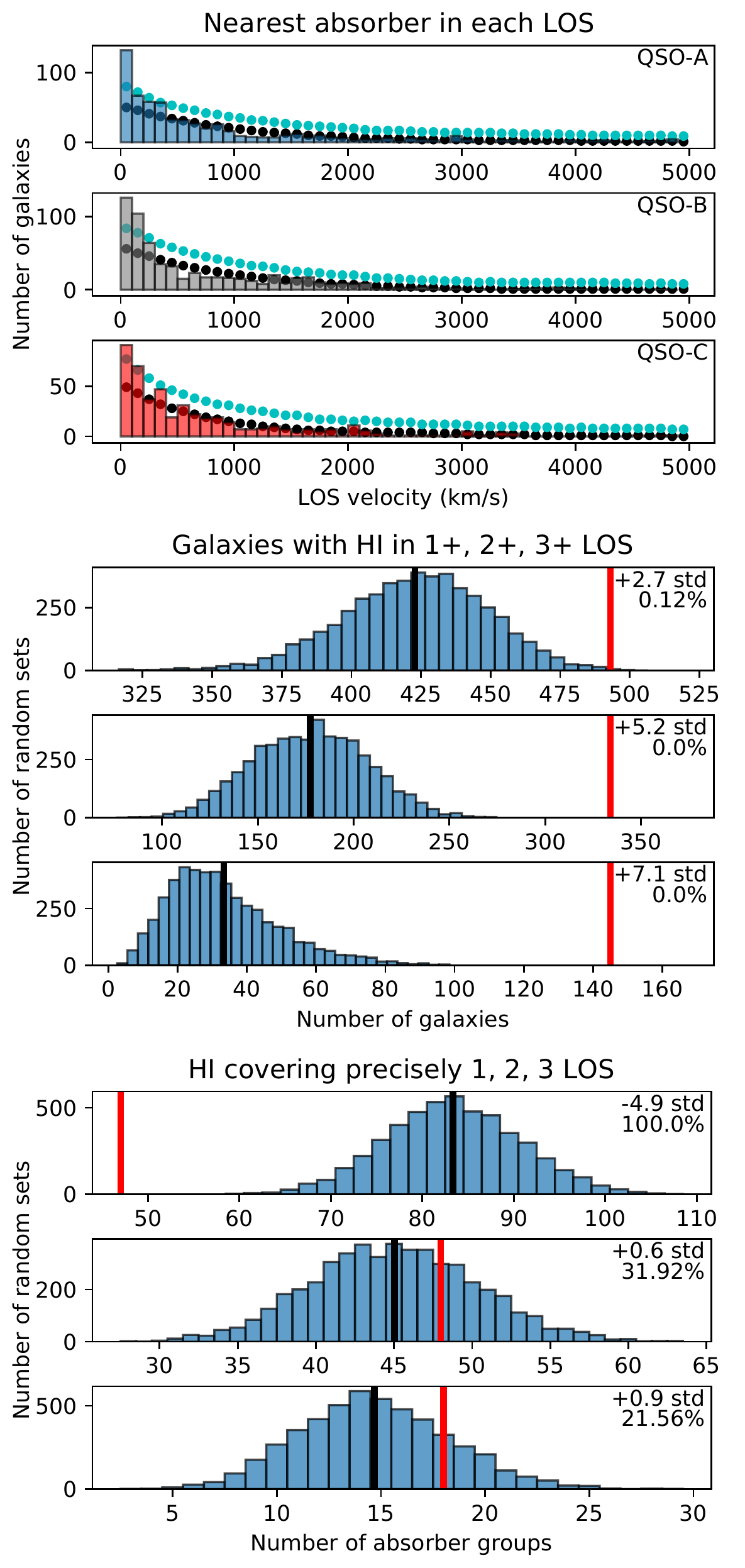}
\caption{Nearest-neighbour and velocity-window groups using the \textbf{full sample}, as described in Section \ref{sec:base}. These compare the absorbers in the real Q0107 sightlines with the ensemble of randomly distributed absorber sets, showing line-of-sight velocity distances between galaxies and absorbers, number of galaxies with associated \ion{H}{i} in one or more lines-of-sight, and number of absorber groups covering one or more lines-of-sight. Features are identical to those in Figure \ref{fig:randoms_coldens} and are described in detail there. \textit{Top 3 panels:} The nearest absorber to each galaxy in each sightline. \textit{Middle 3 panels:} The number of galaxies around which absorption is seen within 500 $\textrm{km}$ $\textrm{s}^{-1}$ in at least one, at least two and all three sightlines.  \textit{Lower 3 panels:} The number of absorber groups covering precisely 1, 2 and 3 sightlines within 500 $\textrm{km}$ $\textrm{s}^{-1}$.  \label{fig:randoms_base_v500}}
\end{figure}

Figure \ref{fig:randoms_base_v500} shows the results from the full sample, in which we consider all observed \ion{H}{i} absorption systems, and all galaxies with `a' and `b' redshift flags within 2.5 Mpc of a sightline. The top panels show the velocity difference between each galaxy and its nearest-neighbour absorber in each sightline. For each sightline, the number of galaxies with absorption within 100 $\textrm{km}$ $\textrm{s}^{-1}$ (the innermost bin) is above the 99th percentile of random absorber sets, and remains consistently above the median out to 400 $\textrm{km}$ $\textrm{s}^{-1}$. This suggests that most of the physical associations between galaxies and absorbers occur with smaller velocity differences. For most of this work, we use a velocity cut of 500 $\textrm{km}$ $\textrm{s}^{-1}$, thus capturing most of the likely galaxy--absorber groups whilst minimizing the noise from unrelated pairs. This is also directly comparable to the grouping used by C10.

The middle panels shows the number of galaxies for which absorption systems are found in at least 1, at least 2, and all three sightlines within the 500 $\textrm{km}$ $\textrm{s}^{-1}$ window. In each case significantly more galaxies have absorption in the real Q0107 field than expected from the systems with randomly generated absorbers. Only six of the 5000 random distributions show as many galaxies with associated absorption, and no set of randoms has as many matches between a galaxy and multiple absorbers as the real Q0107. The significance of the excess also increases with the number of sightlines covered. This is similar to the results of P06 and C10 (their figure 16), in which there is a significant excess of galaxies associated with absorbers on these scales. The larger sample of absorbers and galaxies in this study has allowed a higher confidence level to be reached.

The lower panels show the number of absorber groups covering one, two and three sightlines respectively. As in C10 (the lowest panel of this plot is directly comparable to the middle panel of their figure 7) the number of triple-absorber groups is larger in the real system than in most sets of randoms. However, the excess is less significant here, with $\approx$ 22\% of random sets showing more triples (as compared to $\approx$ 11\% in the C10 results). This may be due to the improved sensitivity to low-column-density absorption. This observed absorption across all three sightlines within a small velocity window supports the idea that the gas is found in structures at least 500-1000 kpc in extent (the distances between the sightlines).

\subsubsection{Column Density} \label{sec:col_dens}

\begin{figure*}

\includegraphics[width=0.998\textwidth]{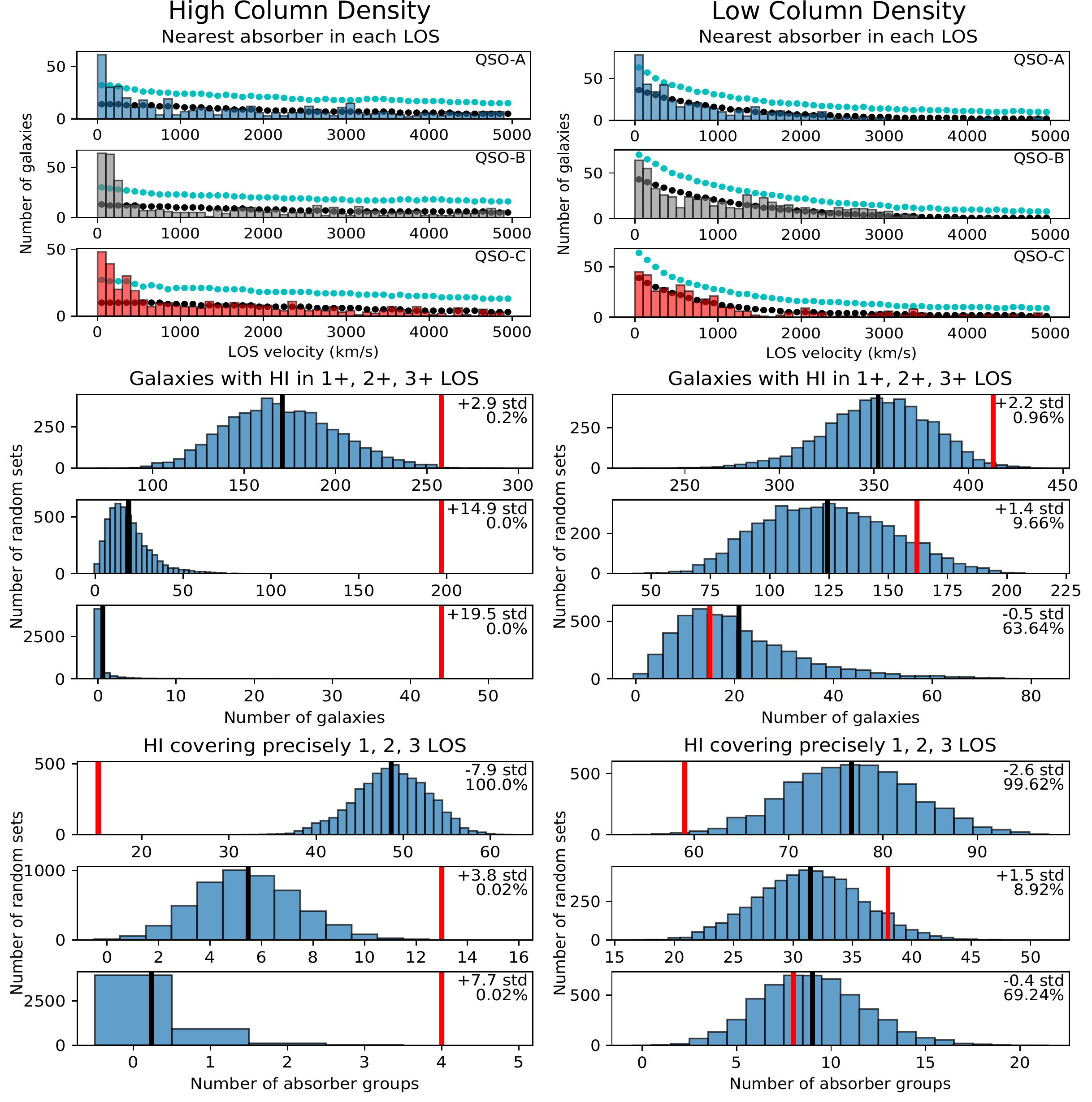}
\caption{Comparison between the real Q0107 system and the ensemble of randomly distributed \ion{H}{i} absorber sets, using the \textbf{column density cut}, as described in Section \ref{sec:col_dens}. We show galaxy and absorber groups involving galaxies with `a' or `b' redshift flags within 2.5 Mpc of at least one line-of-sight, considering only high-column-density (N(\ion{H}{i}) $>$ $10^{14}$ $\textrm{cm}^{-2}$) absorbers in the left column, and lower-column-density absorbers in the right column. The features shown are identical to those in Figures \ref{fig:randoms_base_v500} and \ref{fig:randoms_starform}. \textit{Top 3 panels:} The nearest absorber to each galaxy in each sightline. The histogram shows the velocity difference between each galaxy and the nearest absorber in the sightline given in the top-right of each panel, in 100 $\textrm{km}$ $\textrm{s}^{-1}$ bins. The black and cyan points show the 50th and 99th percentiles of the random distributions respectively, illustrating the excess of galaxies within 300--500 $\textrm{km}$ $\textrm{s}^{-1}$. Note that each galaxy must appear once in each of the three panels, so an excess in the inner bins must be accompanied by a deficit in the bins with higher velocity separations. \textit{Middle 3 panels:} The number of galaxies around which absorption is seen within 500 $\textrm{km}$ $\textrm{s}^{-1}$ in at least one, at least two and all three sightlines. The histograms show the distribution of the random sets, with the black vertical line showing the mean, and the red vertical line shows the number observed in the real Q0107 system. In the top-right we state the number of galaxies with associated \ion{H}{i} found in the real system relative to the mean of the distribution of randomized sets (in units of the standard deviation of the random distribution), and the percentage of random sets for which the number of galaxies with associated \ion{H}{i} absorption is greater than or equal to the number in the observed system. \textit{Lower 3 panels:} The number of absorber groups covering precisely 1, 2 and 3 sightlines within 500 $\textrm{km}$ $\textrm{s}^{-1}$, with features as in the middle three panels. The red vertical line again shows the number of absorber groups covering 1, 2 or 3 sightlines in the observed Q0107 field. This lies further to the left in the upper of these panels for more strongly clustered absorbers, as absorbers are more likely to be coincident with those in another line-of-sight.  \label{fig:randoms_coldens}}
\end{figure*}


We then extend this test by cutting the absorber sample by column density, using a cut of N(\ion{H}{i}) = $10^{14}$ $\textrm{cm}^{-2}$ identical to that used in T14, and perform this analysis on both the high- and low-column density samples. Figure \ref{fig:randoms_coldens} compares the results obtained from these two samples, with the left column giving results from the high-column-density sample, and the right column showing the low-column-density sample.

Using these constraints, the nearest-neighbour results (top) show a much clearer excess of galaxy-absorber pairs in the innermost velocity bins when only high-column-density absorbers are considered, substantially above the 99\% level in all three sightlines. In the low-column-density case, the excess of small-velocity pairs is below the 99\% level in two of the sightlines.

The excess of galaxies with associated absorption is more significant in the high-density sample (middle-left) than the low-column-density sample (middle-right). Indeed, there is no significant excess of galaxies with absorption in multiple sightlines in the low-density case. Similarly, the significance of the excess of real absorber triples is greater in the high-column-density case (bottom panels), and the low-column density observed coincidences are also consistent with the randomly generated sample.

These results confirm previous observations that high-column-density gas preferentially resides close to galaxies, in the CGM or intra-group medium, whereas the excess of low-column-density absorption around galaxies is less significant, suggesting these absorbers occur in the IGM. As only four absorber groups exist with high-column-density absorption in all three lines-of-sight, and these lie at the same redshift as more than 40 galaxies, these triplets correspond to galaxy overdensities.

That low-column-density absorption is not found in multiple sightlines at a significantly higher frequency than in the random distribution may also indicate that these weak absorbers do not often form Mpc-scale structures. 

These results are consistent with those in T14, in which the correlation function between weak absorbers and galaxies suggests that they do not trace the same dark matter distribution, and the low auto-correlation between low-column-density absorbers indicates that they rarely form large structures. \citet{burchett2020} suggest that these column densities should be tracing the outer regions of filaments, as well as some overdensities in voids. This could lead to detection in two sightlines if the filament is aligned across two sightlines, but is unlikely to produce triplets, a possible explanation for the $\sim$ 1.5-$\sigma$ excess of low-column-density, two-sightline detections. 

We also repeated this test after randomly discarding low-column-density absorbers until the sample sizes of high- and low- density absorption features were of equal size. There were no significant differences in the results, confirming that the greater excess seen in the high-column-density case is not merely an effect of the larger sample size. \added[id=AAB]{Due to the higher detection limit in the FOS spectra, there is also a difference in redshift between the low- and high-column-density absorbers, with low-column-density systems rarer at z $\gtrsim$ 0.48 (median redshift of low- and high-column-density absorbers 0.32 and 0.52 respectively). However, the results are similar if we only include absorbers found in the COS spectra. Redshift evolution of the IGM is found to be slow at z $\lesssim$ 1 \citep[e.g][]{kim2020}, so no substantal difference is expected.}

\subsubsection{Star Formation}\label{sec:randoms_sf}

\begin{figure*}

\includegraphics[width=\textwidth]{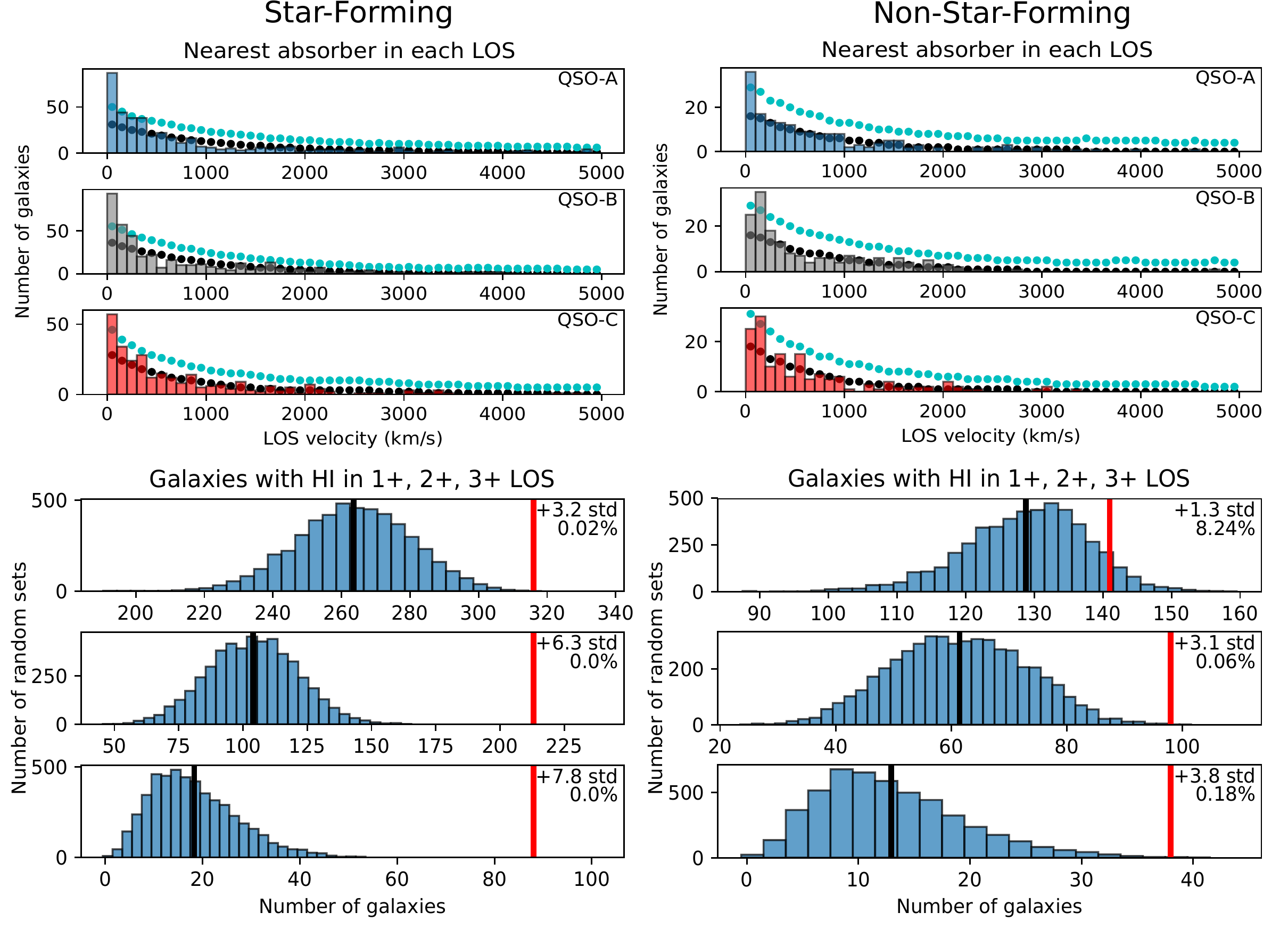}
\caption{Nearest-neighbour and velocity-window groups using the \textbf{star-formation cut}, as described in Section \ref{sec:randoms_sf}. These compare the the real Q0107 system with the ensemble of randomly distributed absorber sets, showing line-of-sight velocity distances between galaxies and absorbers, number of galaxies with associated \ion{H}{i} in one or more lines-of-sight, and number of absorber groups covering one or more lines-of-sight. Features are identical to those in Figure \ref{fig:randoms_coldens} and are described in detail there. \textit{Top 3 panels:} The nearest absorber to each galaxy in each sightline. \textit{Lower 3 panels:} The number of galaxies around which absorption is seen within 500 km/s in at least one, at least two and all three sightlines. The absorber-only groups are omitted as these are not affected by the cut, so the results are the same as in the lower panel of Figure \ref{fig:randoms_base_v500}. \label{fig:randoms_starform}}
\end{figure*}

Another cut was made on the star-formation of galaxies in the sample (using the star-formation class for each galaxy defined in Section \ref{sec:sf_classes}). Figure \ref{fig:randoms_starform} shows these results. The excess of groups in the real system is more significant in the star-forming case (lower-left) than the non-star-forming case (lower-right). This can also be seen in the nearest-neighbour matching (upper-left and upper-right), in which the excess of galaxies in the innermost bins is greater in the star-forming sample. 

These results show absorption is more likely to be detected around star-forming galaxies on scales of hundreds of kpc. This is similar to the results from \citep{chen2009}, in which the correlation between Ly$\alpha$ absorbers and galaxies is stronger among emission galaxies than absorption galaxies. \added[id=AAB]{Our results are generally consistent with those in T14, despite the different approach. They do not find a significant difference in the correlation slope or length between the strong-\ion{H}{i}/SF-galaxy and strong-\ion{H}{i}/non-SF-galaxy cross-correlations. However, they show that the strong/SF results are consistent with linearly tracing the same dark matter distribution, whilst the strong/non-SF results are not, due to the greater auto-correlation between non-SF galaxies. Therefore the probability of finding a strong \ion{H}{i} absorber near a non-SF galaxy is lower than that of finding a strong \ion{H}{i} absorber near a SF galaxy, in agreement with our results.}

\added[id=AAB]{Whilst the star-forming galaxies in our sample are more likely to be surrounded by detectable levels of \ion{H}{i} gas, it is difficult to determine the cause of this. We note that excess absorption within the virial radius is not sufficient to explain the excess of two- and three-sightline absorption, as the distance between the sightlines is larger than the virial radius for most of our galaxies.}

\added[id=AAB]{There are suggestions from simulations that stellar-feedback-driven outflows may extend to and beyond the virial radius \citep[e.g.][]{defelippis2020, mitchell2020a, hopkins2021}, so their presence around star-forming galaxies could lead to the observed absorption.} If outflows make a substantial contribution to this excess absorption, we may expect to see a broader distribution of galaxy-absorber velocity offsets in the SF sample, arising from the line-of-sight component of the outflow velocity \citep[see e.g.][]{chen2020}. This broader distribution is not seen in our data, which would suggest that outflows do not make a large contribution on these scales. However, there are potential redshift errors in the VIMOS galaxies in the same 100--200 $\textrm{km}$ $\textrm{s}^{-1}$ regime as the likely outflow velocities (see Table \ref{table:z_errors}), which could similarly broaden the distribution. We further discuss the presence of outflows in Section \ref{sec:orientations}. 



\added[id=AAB]{We consider the environments in which the SF and non-SF galaxies are likely to be found.} Non-SF galaxies are more likely to reside in groups (41\% of non-SF and 32\% of SF galaxies lie in groups of five or more galaxies, using the friends-of-friends algorithm described in Appendix \ref{sec:fof_groups}), as expected due to quenching when galaxies fall into larger haloes \citep[e.g.][]{wetzel2013}. As these groups will often correspond to larger overdensities in the cosmic web than single galaxies, we may expect increased incidence of \ion{H}{i} absorber groups around non-SF galaxies, the opposite effect to that observed. However, the higher virial temperatures of group haloes would lead to a suppression of neutral hydrogen, possibly negating this effect. 


\added[id=AAB]{We also consider the effects of biases in our sample selection (as discussed in Section \ref{sec:sf_classes}). Firstly, to remove any effects of having a larger star-forming than non-star-forming sample, the analysis was rerun after randomly discarding star-forming galaxies until the sample sizes were identical, and the significance of the excess in galaxy-absorber groups was only marginally reduced in each case. Whilst there are small biases in the mass and impact parameter distributions of the two samples, a Kolmogorov-Smirnov (K-S) test cannot distinguish them at the 5\% level, and they appear localised to the mass tails and the smallest impact parameters (i.e. the MUSE fields). Removing the mass tails and/or MUSE-only galaxies again slightly reduces the significance of the excess, but does not substantially affect the comparison. } 




\added[id=AAB]{The bias of SF galaxies towards higher redshifts does appear to have an effect, but this is found to be }  primarily due to the absorber selection function removing lower column density absorbers at higher redshifts (as seen in Figure \ref{fig:ew_detection}). If only high-column-density absorbers are considered, no substantial difference between high-z and low-z coherence is found, but there remains a larger excess of absorption around star-forming galaxies than non-star-forming galaxies. 


\added[id=AAB]{The difference between our SF and non-SF samples is therefore unlikely to have arisen from any sample biases, and instead does indicate that star-forming galaxies are more likely than quiescent galaxies to exhibit \ion{H}{i} absorption covering single and multiple lines-of-sight.}


\section{Galaxy Orientations}\label{sec:orientations}

As described in Section \ref{sec:gal_data} and shown in Figure \ref{fig:survey_layout}, we have obtained MUSE data for the $1'\times 1'$ fields around QSOs A and B, in addition to high-resolution HST imaging of a larger field (but not extending as far as QSO-C). We used GALFIT \citep{peng2002} to determine \added[id=AAB]{the} position angles \added[id=AAB]{of galaxies in the HST field,} as described in Section \ref{sec:HST}. This allows galaxy--quasar pairs in which absorption is detected (and those where it is not) to be shown as a function of position angle and impact parameter. We can then determine whether absorption lies preferentially along the major or minor axis of the galaxy, and identify possible co-rotating material and polar outflows.

The MEGAFLOW survey mostly covered impact parameters out to $\sim$ 100 kpc using \ion{Mg}{ii}, focusing their selection on sightlines with high-equivalent-width absorbers in their SDSS spectra \citep{schroetter2016}. They find a clear bimodality in position angle, with more absorbers found along both the major and minor axes, identified with discs and outflows respectively. It must be noted that they remove absorbers found with three or more nearby galaxies, and assign a single galaxy to the absorber in cases with two nearby galaxies.

Similarly, \citet{bordoloi2011} and \citet{bordoloi2014a} look for \ion{Mg}{ii} around zCOSMOS galaxies, finding strong minor-axis absorption attributed to outflows that is much stronger than the major-axis absorption, and an increasing outflow equivalent width with host galaxy SFR. They only find this for impact parameters $\lesssim$ 50 kpc, but also include group galaxies in their analysis, obtaining results consistent with a superposition of outflows from group members.

On the other hand, \citet{dutta2020} find that their sample, primarily consisting of \ion{Mg}{ii} absorption at larger impact parameters, does not show such a bimodality \citep[a similar result is found by][]{huang2021}, nor can their measurements in group environments be fully explained by a superposition of absorption from individual group members. This suggests that, at least in \ion{Mg}{ii}, the disk/outflow dichotomy does not extend to scales much beyond $\sim$ 50 kpc. 

\added[id=AAB]{This appears in contrast to some recent models. For example, \citet{hopkins2021} find that their simulations (based on the FIRE simulations, with the addition of cosmic ray effects) can allow biconical outflows to reach megaparsec scales. Outflows in the EAGLE simulations are also seen to maintain their bi-directional structure to at least the virial radius of $\sim 10^{12}$ $\textrm{M}_{\odot}$ halos \citep{mitchell2020a}, whilst Illustris also features hot outflowing material and cool co-rotating material along the minor and major axes out to close to the virial radius \citep{defelippis2020}.}

Using \ion{H}{i} absorption from lines-of-sight with no pre-selection allows an unbiased sample to extend to larger impact parameters \added[id=AAB]{than most \ion{Mg}{ii} studies}.
\added[id=AAB]{When} considering galaxies in groups\added[id=AAB]{, we avoid} 
artificially selecting the associated galaxy to each absorber \added[id=AAB]{, instead including all galaxy--absorber pairs within the 500 $\textrm{km}$ $\textrm{s}^{-1}$ cut}. We test for this bimodality in position angle for the galaxy--absorber pairs in our sample using the Hartigan dip test \citep{hartigan1985}, which calculates the likelihood of observing the `dip' in the sample histogram if the underlying distribution is unimodal. As in Section \ref{sec:coherence}, we also compare the results when the sample is split into complementary subsamples such as high- and low-column-density absorbers, and star-forming and non-star-forming galaxies. 

\subsection{Hydrogen}

The results from applying this test to the distribution of galaxy--absorber pairs involving \ion{H}{i} absorption are summarized in Table \ref{table:pa_results}. Where the sizes of the complementary samples (paired using horizontal lines in the table) are substantially different, we randomly discard detections from the larger sample until the sizes match in order to perform a fair comparison. We repeat with 100 random samples and take a median, giving the results in brackets.

\subsubsection{\added[id=AAB]{Full Sample}}

\begin{figure}
\includegraphics[width=\columnwidth]{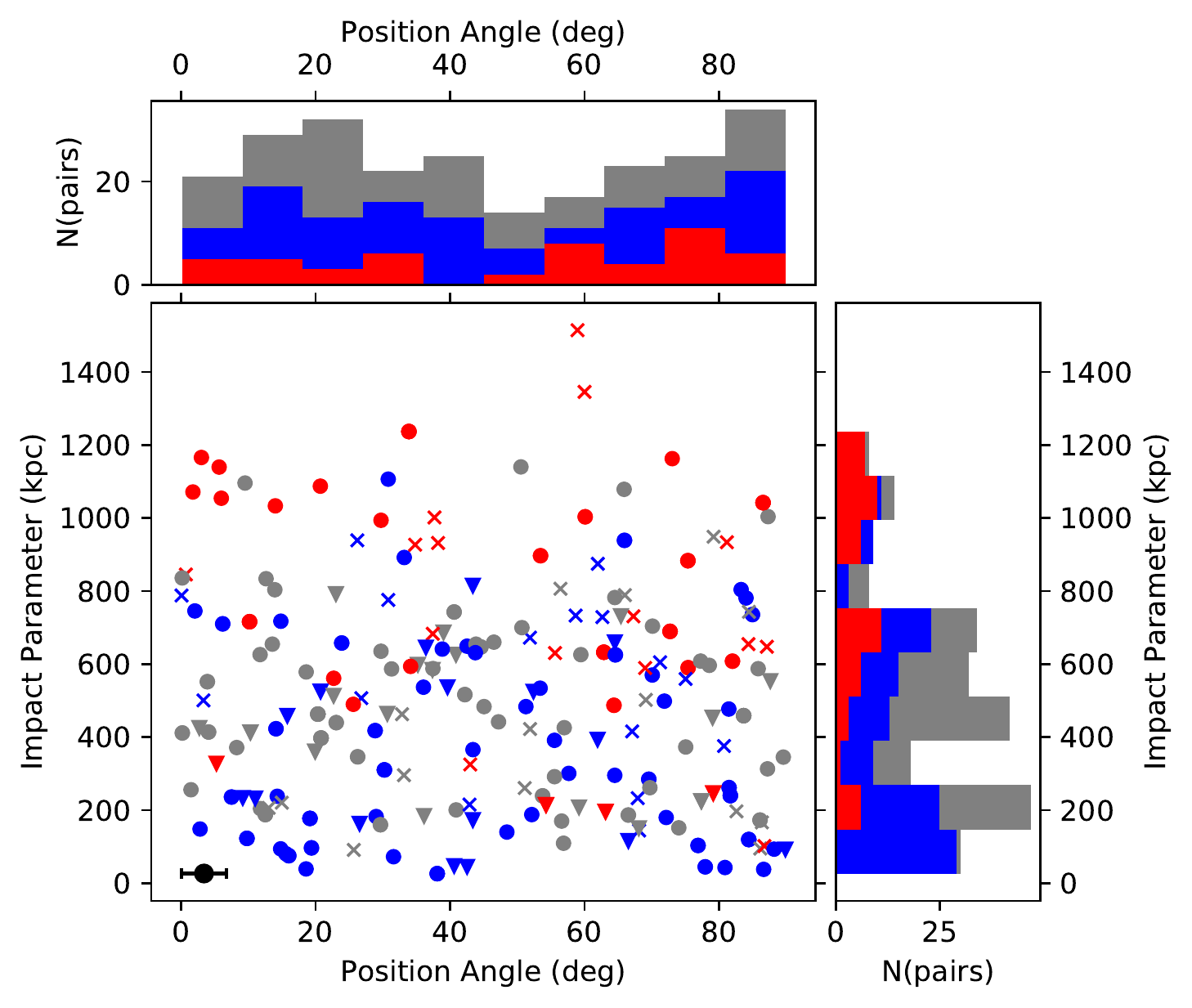}
\caption{The position angle and impact parameter distributions of galaxy-absorber pairs involving \ion{H}{i} absorption features around all galaxies in the sample for which position angles were measured. Position angles close to  $0^{\circ}$ represent absorption found close to the projected galaxy major axis, whilst those close to $90^{\circ}$ show absorption found near the minor axis. Points are coloured by quasar sightline (A, B and C shown in blue, grey and red respectively), using triangles for galaxies in groups of five or more galaxies and circles for non-group galaxies, and non-detections are marked with crosses. \added[id=AAB]{Only the detected absorbers are included in the histograms.} The bars on the black point in the lower-left are illustrative of the median error in each axis. \label{fig:pa_imp_all}}
\end{figure}

Figure \ref{fig:pa_imp_all} shows the position angle against impact parameter for the full sample of galaxies and absorbers, associating each galaxy with all absorbers within 500 $\textrm{km}$ $\textrm{s}^{-1}$. \added[id=AAB]{Non-detections are shown on the scatter plot, but only the position angles and impact parameters of detected absorption are included in the histograms.} There is a visible bimodality, and the dip test finds a significant result, returning a p-value of 0.025. This reproduces in \ion{H}{i} the bimodality obtained using \ion{Mg}{ii} in \citet{zabl2019} and \citet{martin2019a}, although extending to larger impact parameters, suggesting that some fraction of our observed \ion{H}{i} is tracing the same outflowing and accreting material. We note that the major axis absorption we observe covers position angles $\lesssim 40^{\circ}$. \added[id=AAB]{It is not simple to distinguish between a thin disk viewed at moderate inclination angles, and a thicker `wedge' of material.}


Some studies \citep[e.g][]{tempel2013, zhang2015} suggest that galaxy spins are preferentially aligned with or perpendicular to the surrounding large-scale structure. Observing the cosmic web around these galaxies, as traced by \ion{H}{i}, could produce a bimodality. This alignment is a weak effect (in \citealt{tempel2013}, galaxies are a maximum of $\approx$ 20\% more likely to be aligned over a random distribution), so unlikely to fully explain our stronger observed bimodality. 

\added[id=AAB]{The overall position angle distributions of galaxies and galaxy-sightline pairs are consistent with uniform, with neither the dip test nor a K-S test against a uniform distribution showing a significant result. Therefore our bimodality is not due to an inherent alignment between galaxies in our sample. It is instead most likely due to the inflow/outflow dichotomy discussed above.}

\subsubsection{\added[id=AAB]{Galaxy Groups}}

Cutting the sample to only include galaxies in groups of five or more decreases the significance of the bimodality to $\approx$ 10\%, whereas no bimodality is seen among the non-group galaxies (using the friends-of-friends algorithm described in Appendix \ref{sec:fof_groups}). 

\added[id=AAB]{The significance of the bimodality is expected to be reduced in galaxy groups. If the gas is primarily a superposition of outflows and/or accretion from individual galaxies, the signal would still be partially masked by other galaxies being paired with the same absorber. If the gas does not form these structures within most groups, and instead forms an intra-group medium not attributable to a single galaxy, this bimodality should not be visible.} 

\added[id=AAB]{The bimodality we see in the in-group sample is not significant at the 5\% level, and}
the significance is further reduced
when galaxy-absorber pairs are randomly discarded until the 
\added[id=AAB]{group and non-group} samples contain the same number of pairs. Our results are consistent with the difference in the p-values shown in Table \ref{table:pa_results} being primarily due to sample size, as neither sample shows a significant bimodality at the 5\% level\added[id=AAB]{, yet the combined sample does}. A K-S test also fails to find a significant difference between the position angle distributions of galaxy-absorber pairs of these two samples. 

No significant effect on the bimodality or the position angle distribution is found when using shorter linking lengths, when adjusting the minimum number of galaxies needed to constitute a group between three and five \added[id=AAB]{, or when splitting the non-group galaxies into those with no detected neighbours and those with 1-3 neighbours}.

\subsubsection{\added[id=AAB]{Impact Parameters}}

When the sample is cut to absorber-galaxy pairs with an impact parameter of less than 500 kpc, the bimodality remains strong, a Hartigan dip test returning a significance of 3\%, whereas the pairs with an impact parameter larger than 500 kpc exhibit no significant bimodality. The difference between the large- and small- impact parameter results may suggest that the model of minor axis outflow and major axis accretion can extend well beyond the $\sim$ 100 kpc observed in the MEGAFLOW results \citep{schroetter2016}. We confirm that this bimodality is not entirely driven by galaxy-absorber pairs on small scales, obtaining a significant result from pairs with impact parameters of 200--500 kpc.

\begin{figure}
\includegraphics[width=\columnwidth]{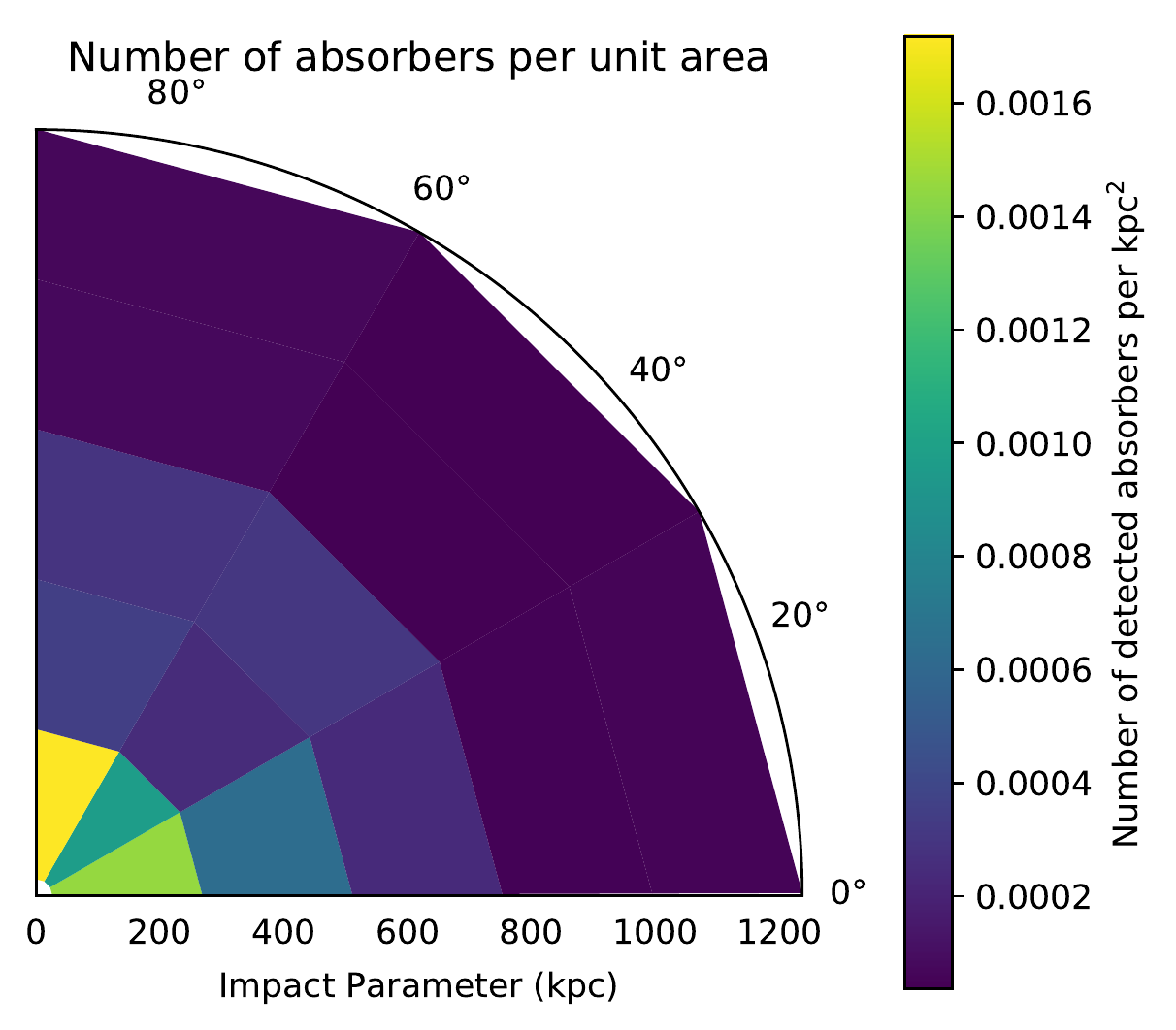}
\caption{The position angles and impact parameters of \ion{H}{i} absorbers detected around galaxies, shown in bins of $\sim$ 250 kpc $\times$ $30^{\circ}$. In this representation the galaxy major axis lies along the x-axis. Colour scale shows number of detections per projected $\textrm{kpc}^{2}$. Note that the radial distribution is determined primarily by the geometry of the survey: with QSOs A and B lying near the centre of the HST field, and QSO-C outside the field, only absorbers in the line-of-sight to QSO-C can be found beyond 800 kpc.  \label{fig:polar_area_all}}
\end{figure}

This is further illustrated in Figure \ref{fig:polar_area_all}, in which the galaxy-absorber pairs are binned by position angle and impact parameter. For the two innermost bins in impact parameter ($<$ 500 kpc), the 30-$60^{\circ}$ bin has a clear lack of absorption relative to the major axis and minor axis bins. The third radial bin ($\approx$ 500-750 kpc) does not show this bimodality, suggesting that IGM absorption or other structures such as group material form the dominant component at this scale. Beyond this distance, the sample size is very small. We note that few conclusions can be drawn from the radial distribution, as this depends primarily on the geometry of the sightlines and redshift surveys. (The apparent `edge' at $\approx$ 750 kpc is an artifact of this layout, as this is roughly the maximum distance from the A and B sightlines to the edge of the HST imaging. Most galaxy--absorber pairs beyond this distance involve absorption in sightline C.)

\begin{table}
\begin{center}
\caption{Summary of constraints applied to the sample and resulting bimodalities in the position angle distributions of galaxy--\ion{H}{i} absorber pairs. Columns show: (1) The subsample used, (2) the number of possible galaxy--sightline pairs, accounting for the reduced redshift range for which QSO-C can be observed as well as the additional constraints, (3, 4) the number of galaxy--sightline pairs with and without observed \ion{H}{i} absorption within $\Delta v < 500$ $\textrm{km}$ $\textrm{s}^{-1}$, and (5) the p-value obtained from a Hartigan dip test applied to the position angle distribution of galaxy--absorber pairs with detected absorption. Where the sizes of complementary samples are substantially different, we randomly discard detected galaxy-absorber pairs from the larger sample until the sizes match, repeating 100 times and showing the average result in brackets.}
\label{table:pa_results}
\begin{tabular}{| c| r| r| r| c| }
\hline 
Constraint & Pairs & Det & Non-Det & P-value \\
(1) & (2) & (3) & (4) & (5) \\\hline 

Full Sample & 289 & 242 & 47 & 0.025 \\ \hline
In-Group & 190 & 173 & 17 & 0.091(0.281) \\
Non-Group & 99 & 69 & 30 & 0.511 \\ \hline
r $<$ 500 kpc & 154 & 137 & 17 & 0.031 \\
r $>$ 500 kpc & 152 & 105 & 47 & 0.185 \\ \hline
r $<$ 300 kpc & 93 & 81 & 12 & 0.030 \\
r $>$ 300 kpc & 208 & 161 & 47 & 0.399(0.393)  \\ \hline
200 $<$ r $<$ 500 kpc & 101 & 84 & 17 & 0.033 \\\hline
High-N(\ion{H}{i}) & 245 & 128 & 117 & 0.009  \\
Low-N(\ion{H}{i}) & 229 & 114 & 115 & 0.709 \\ \hline
Star-forming & 187 & 155 & 32 & 0.123(0.224) \\
Non-SF & 83 & 73 & 10 & 0.284 \\ \hline


\end{tabular}

\end{center}
\end{table}

\subsubsection{\added[id=AAB]{Column Densities}}

\begin{figure}
\includegraphics[width=\columnwidth]{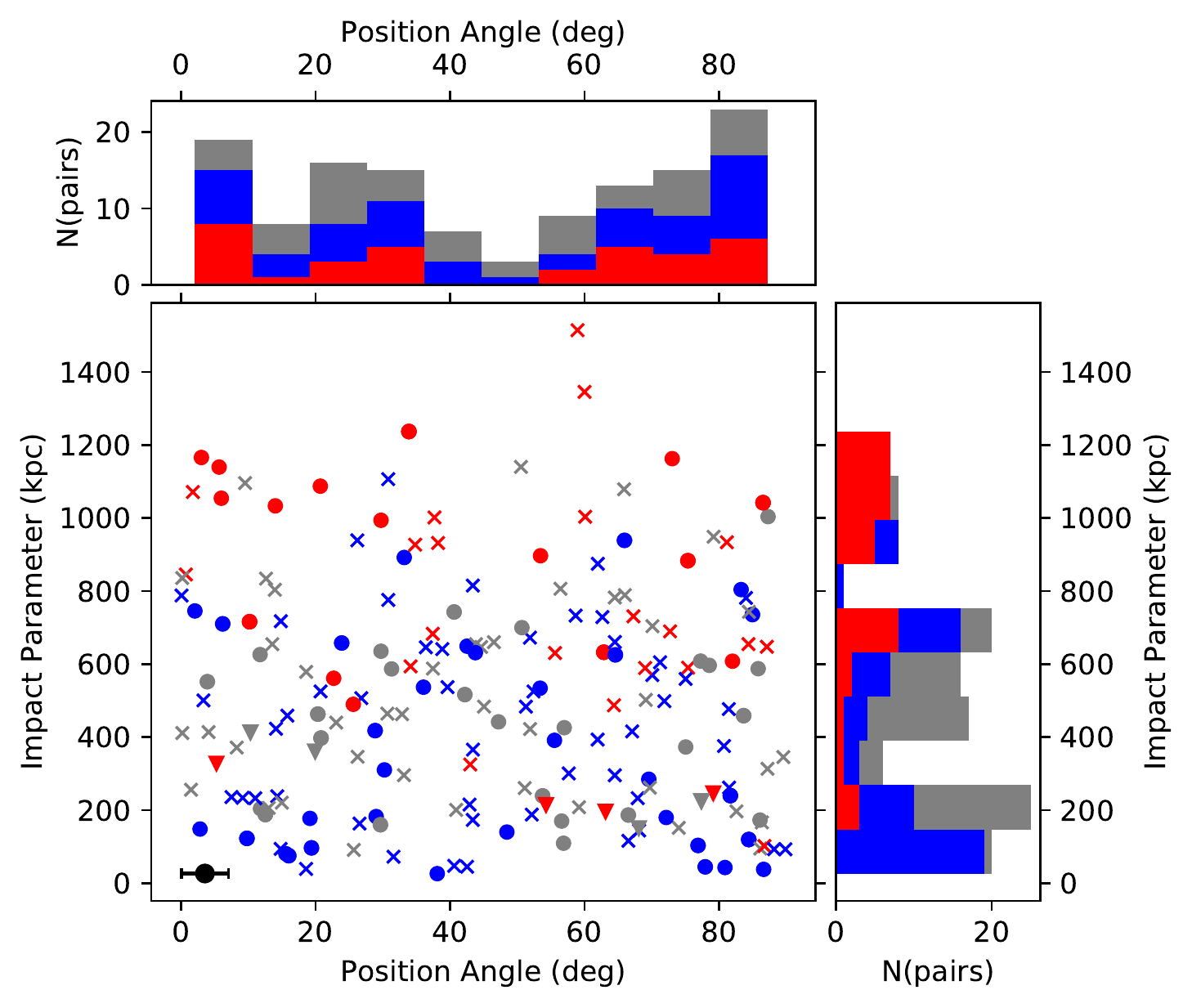}
\caption{The position angle and impact parameter distributions of galaxy-absorber pairs involving \ion{H}{i} absorption features with column densities N(\ion{H}{i}) $>$ $10^{14}$ $\textrm{cm}^{-2}$. Features are as in Figure \ref{fig:pa_imp_all}. \label{fig:pa_imp_highdens}}
\end{figure}

Cutting to column densities N(\ion{H}{i}) $>$ $10^{14}$ $\textrm{cm}^{-2}$ improves the significance of the bimodality to better than 1\%, as shown in Figure \ref{fig:pa_imp_highdens}, whereas no clear bimodality is found for the low-column-density absorbers. These two subsamples show the strongest and weakest bimodalities according to the dip test results, suggesting that high-column-density gas is preferentially found in these putative inflows and outflows. This likely captures much of the same physics as the variation with impact parameter, as high-column-density absorbers are generally found closer to galaxies, for example in \citet{chen2012} and \citet{keeney2017}. \citet{wilde2021} find that the probability of finding \ion{H}{i} above our column density threshold around a galaxy drops to 50\% at impact parameters of $\approx$ 300 kpc, similar to the extent of our bimodality.

\added[id=AAB]{As discussed in Section \ref{sec:col_dens}, the high- and low-column density samples have different redshift distributions, but this does not have a substantial impact on the results. We confirm that the bimodality is retained for the high-column-density absorbers in the COS gratings, removing the FOS absorbers at higher redshifts where low-column-density absorbers are not detected (p=0.018).}

\subsubsection{\added[id=AAB]{Star Formation}}

We also consider the bimodality around star-forming and non-star-forming galaxies.
\added[id=AAB]{Neither subsample shows a significant bimodality}
in position angle of absorption (p-values 0.12 and 0.28 respectively). 
This appears to be primarily due to the sample sizes. When star-forming galaxies are randomly discarded from the sample until the number of detections \added[id=AAB]{around} the star-forming and non-star-forming galaxies are equal, 
\added[id=AAB]{the resulting p-values are similar between the two samples.} A K-S test also fails to find a significant difference between the position angle distributions of absorbers around star-forming and non-star-forming galaxies.

It is expected that more strongly star-forming galaxies are likely to have stronger outflows \citep[e.g.][]{mitchell2020a}. Our star-forming classification may be including many galaxies with SFRs too small to launch large-scale outflows, thus reducing the strength of the observed bimodality. We attempt to test this by applying a cut in sSFR (using the estimates described in Section \ref{sec:sf_classes}) instead of our binary classification. This results in a bimodality significant at the 5\% level in the strongly star-forming sample when using a threshold between $\approx$ 0.05 and 0.1 $\textrm{Gyr}^{-1}$. Higher thresholds leave the star-forming sample too small to obtain a significant result, and lower thresholds give similar results to our original classification. 

This can be discussed in the context of the star-formation comparison in Section \ref{sec:randoms_sf}. Whilst our sSFR estimates have high uncertainties, our finding that more strongly star-forming galaxies show a stronger bimodality is likely an indication that stellar-feedback-driven outflows are a contributor to this bimodality we observe in the position angle distribution of \ion{H}{i} absorbers around galaxies on scales $\lesssim$ 300 kpc. Our result that outflows on these scales are not evidenced by a bimodality around our more inclusive sample of star-forming galaxies, yet larger-scale coherent structures are observed (on the 500-1200 kpc scales probed by coincidences between the lines-of-sight), suggests that these larger-scale structures are not primarily a result of outflows, but are instead a consequence of the environment around these galaxies.


\subsubsection{\added[id=AAB]{Inclination}}

We note that when an inclination cut of $i > 40^{\circ}$ (identical to that used in the MEGAFLOW survey, \added[id=AAB]{\citealt{zabl2019}}) is used to remove face-on galaxies for which the major and minor axes cannot easily be distinguished, the same subsamples show significant bimodalities at the 5\% level as those in Table \ref{table:pa_results}. \added[id=AAB]{For randomly oriented galaxies, cos(i) is expected to be uniform, so the distribution of inclinations is not uniform, but is instead suppressed at low inclinations (face-on galaxies). This leads to only 8 of the 72 galaxies considered throughout this section having inclinations less than $40^{\circ}$, so it is unsurprising that the results do not change.} 
\added[id=AAB]{We also note that a small number of galaxies, although fit well by the GALFIT modelling, have large uncertainties on their position angle (5 have position angle uncertainties $> 20^{\circ}$). Excluding these galaxies also has no effect on which subsamples show bimodalities significant at the 5\% level. }


\added[id=AAB]{We briefly consider variation with inclination, by dividing the $i > 40^{\circ}$ sample into two bins of 40-65 and 65-$90^{\circ}$ (which contain 39 and 25 galaxies respectively). Neither sample shows a clear bimodality, and both are consistent with the overall distribution using the K-S test. Interestingly, the two bins are not consistent with each other (p $\approx$ 0.04). This appears to be mostly driven by galaxy--absorber pairs with impact parameters in the 250-500 kpc bin.} 

\added[id=AAB]{We find that 26 of the 39 pairs in this impact parameter range involving `intermediate-inclination' galaxies lie in the 0-$30^{\circ}$ major axis bin in position angle, whilst 14 of the 23 pairs involving $i > 65^{\circ}$ galaxies lie in the 60-$90^{\circ}$ minor axis bin in position angle. Whilst these sample sizes are relatively small, this perhaps supports the presence of a disk-like structure along the major axis with a small cross-section when viewed close to edge-on.}

\subsubsection{\added[id=AAB]{Closest Galaxies}}

\added[id=AAB]{We note that if we reduce our sample to only the closest galaxy in impact parameter to each absorber (within the $500$ $\textrm{km}$ $\textrm{s}^{-1}$ cut), the bimodality becomes somewhat stronger, with significance improving from $\approx2.5\%$ to $\approx1.0\%$. This will capture the physical origin of the absorbing material in many cases, but does not appear to in all cases, as the galaxy-absorber pairs removed do themselves show some hint of bimodality (with a significance of $\approx7\%$). Splitting the sample of galaxy-absorber pairs involving only the closest galaxy produces results similar to those in Table \ref{table:pa_results}, with no sub-sample crossing the 5\% threshold. We do not take this approach through our earlier analysis as the cut appears to remove some physical associations and substantially reduces the sample size in some cases. This makes detecting a bimodality more difficult, especially in the in-group, SF and non-SF sub-samples.}

\subsection{Metals} \label{sec:metals}

We briefly discuss here the presence of metals in our absorption-line sample. Lines in the spectra exist that are identified with numerous ions, including \ion{C}{i}-\ion{C}{iv}, \ion{Si}{ii}, \ion{Si}{iii}, \ion{O}{i}-\ion{O}{iv} and \ion{O}{vi}. Only \ion{O}{vi} forms a significant sample, with 34 lines identified across the three spectra.

\begin{figure}
\includegraphics[width=\columnwidth]{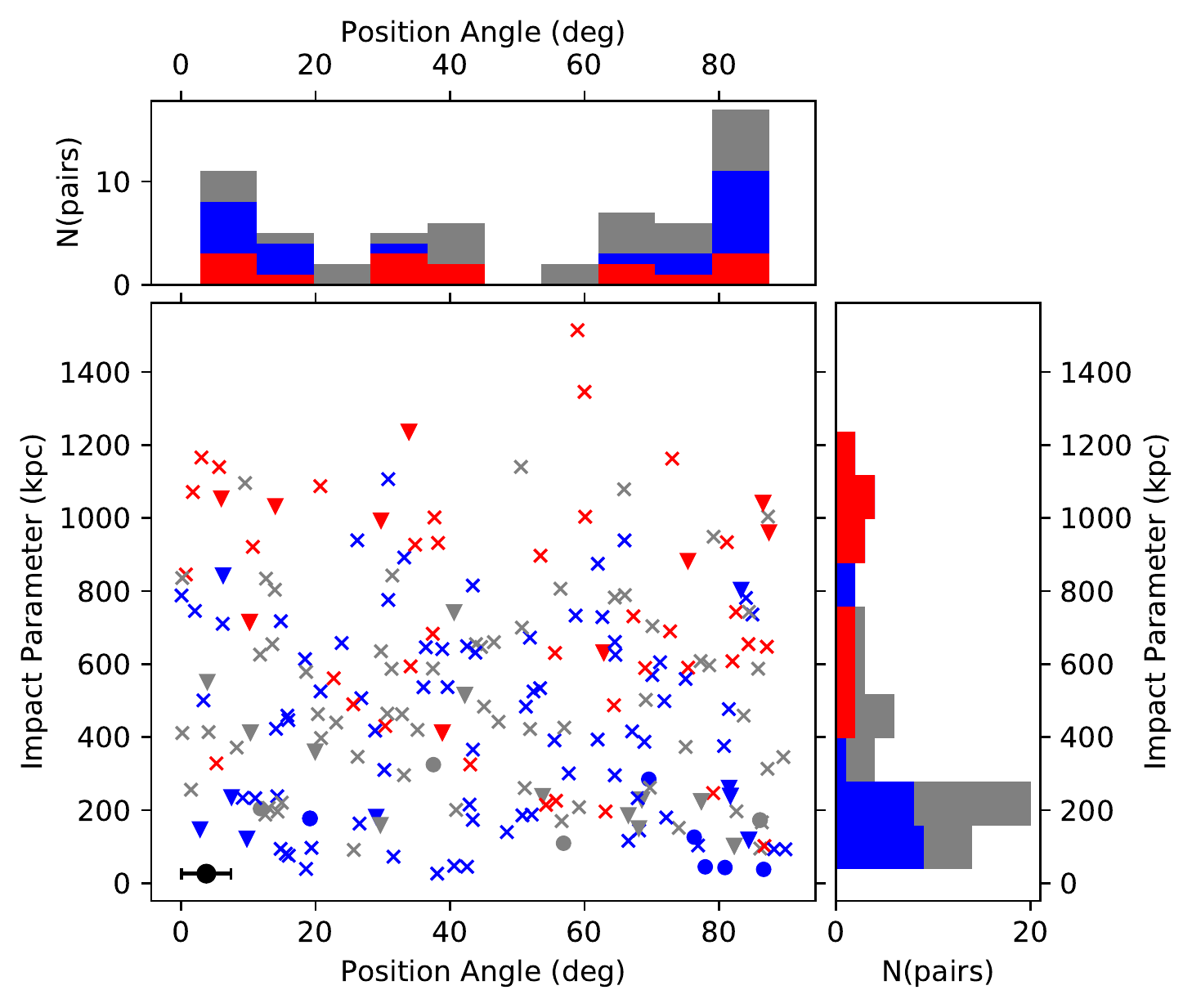}
\caption{The position angles and impact parameters of detected \ion{O}{vi} absorption features around all galaxies with measured position angles. Features shown are as in Figure \ref{fig:pa_imp_all}. \label{fig:pa_imp_ovi}}
\end{figure}

Figure \ref{fig:pa_imp_ovi} shows the resulting distribution of position angles and impact parameters. The bimodality is again visible by eye, and statistically significant (p $\sim$ 0.009 from the Hartigan dip test). It must be noted that all non-group detections occur at impact parameters of less than 350 kpc \added[id=AAB]{(postage-stamp images of the 6 non-group galaxies with detected \ion{O}{vi} absorption are shown in Appendix \ref{sec:galfit_results})}. It may also be suggested that the minor-axis `peak' is stronger relative to the major-axis than in the \ion{H}{i} figures. This is in general agreement with the results from \citet{kacprzak2015}, in which a bimodality is also found around isolated galaxies, with higher \ion{O}{vi} equivalent widths along the minor axis. 


We also compare the column-density distributions of \ion{O}{vi} detections in galaxy groups with those not in groups. A K-S test indicates that these distributions are likely different (98\% confidence), with a substantial number of in-group absorbers observed with lower column density. 

This is again consistent with the disk/outflow model at small scales, whereas all metal absorbers found at large impact parameter from galaxies occur in groups. This could be due to absorber-galaxy pairs involving an absorber associated with a different (possibly undetected) galaxy in the same group, virialized gas within the group halo, or due to tidal and intra-group material within the group. Whilst observations of this material in emission \citep[e.g][]{fossati2019a} allow us insight into the structure of this material on small scales, the lower column densities found on larger scales cannot easily be probed in this way. \citet{morris1994} found that tidal absorbers extending to 1 Mpc radius would be consistent with observations. Unfortunately, this is approximately the limit of impact parameters for which position angles are available, so we cannot test whether this \ion{O}{vi} extends to larger radii.

\citet{oppenheimer2016} predict that the presence of \ion{O}{vi} should be suppressed in group haloes due to a higher virial temperature than $L_{\star}$ galaxy haloes, reducing the fraction of \ion{O}{vi} in favour of higher ions. A similar suppression is found in \citet{wijers2020a} for high-mass haloes.

Alternatively, observations and radiative transfer modelling from \citet{werk2014} suggest that the warm gas phase is not supported hydrostatically in $L_{\star}$ galaxies, favouring a galactic `fountain' over a virialized halo. The strong bimodality we observe in our data favours the galactic fountain for non-group galaxies. That the scale of \ion{O}{vi} in non-group galaxies is similar to the extent of the observed bimodality in \ion{H}{i} also supports the existence of such fountains extending to $\sim$300 kpc. The apparent shift in the ratio between the major- and minor-axis peaks, towards the minor axis, further fits this picture, with warmer, higher-metallicity outflows exhibiting more \ion{O}{vi} than cool accreting flows.

However, our sample contains relatively few isolated $L_{\star}$ galaxies (see Figure \ref{fig:lum_z}), with most substantially smaller or lying in groups of galaxies. Either case moves the virial temperature of the halo away from the $\sim 10^{5.5}$K at which \ion{O}{vi} is most prominent, making these virialized haloes harder to observe. Therefore, whilst a basic model without such haloes is consistent with the data, producing \ion{O}{vi} from fountains and group interactions, we cannot rule out their presence. With our small number of isolated $L_{\star}$ galaxies, we cannot detect any `peak' in absorber number or column density around $L_{\star}$ galaxies.

We also note that there are four redshifts at which \ion{O}{vi} absorption occurs in two sightlines within 500 $\textrm{km}$ $\textrm{s}^{-1}$, with no three-sightline \ion{O}{vi} systems. In one of these systems, both sightlines featuring \ion{O}{vi} lie within $30^{\circ}$ and 200 kpc of the minor axis of an isolated 1.5 $L_{\star}$ galaxy. The other three systems exist in galaxy groups, where it is difficult to distinguish between tidal debris, virialized haloes, and warm outflows. More detailed ionization modelling of the absorption in two of these groups are described in \citet{muzahid2014} and \citet{anshul2021}.

\subsection{Comparison between major and minor axes}

We now compare the properties of absorption along the major- and minor- axes. We take two samples of galaxy-absorber pairs, with position angles $< 40^{\circ}$ and $> 50^{\circ}$ respectively. We remove face-on galaxies by requiring a galaxy inclination of $> 40^{\circ}$ and impose a 500 kpc maximum impact parameter to focus on the regions in which the bimodality is significant\footnote{These cuts in inclination and azimuthal angle cover an identical region to the MEGAFLOW survey for the major axis sample \citep{zabl2019}, whilst the slightly larger region along the minor axis is used both for symmetry and to produce a sample of identical size. The major axis cut is also identical to that shown in Figure \ref{fig:inc_imp_kinematics}, although we do not require observed kinematic data in this case.}. This returns 67 galaxy--\ion{H}{i} pairs in each sample.

A K-S test reveals no difference between the column density or Doppler width distributions of these two samples. We also compare the galaxy--absorber velocity offsets around moderately inclined and edge-on galaxies ($40-65^{\circ}$ and $> 65^{\circ}$ inclination respectively), finding no significant difference. If our observed position angle distribution is primarily due to major axis co-rotation and minor axis outflows, we may expect to see a wider velocity distribution along the minor axis of moderately inclined galaxies and the major axis of edge-on galaxies, due to the larger line-of-sight velocity components. This is not observed in our data, although the further split results in a small sample size.


However, 14 of the major-axis absorbers have associated \ion{O}{vi}, whilst 28 of the minor-axis absorbers show this association. A similar ratio is found when using a 300 kpc maximum impact parameter. This difference in \ion{O}{vi} incidence is found in \citet{kacprzak2015} and \citet{kacprzak2019}, but is attributed to a higher \ion{H}{i} column density rather than metallicity differences. We do not find any significant increase in \ion{H}{i} column density along the minor-axis, so the increased \ion{O}{vi} found near the minor-axis may be due to more warm-hot material than in the absorbers found near the major axis.




\section{Kinematics}\label{sec:kinematics}

We attempt to determine whether the major-axis population discussed above is indeed the same co-rotating material identified in \ion{Mg}{ii} \citep[e.g][]{zabl2019}, as well as found in simulations \citep[e.g][]{huscher2021, defelippis2020}. We compare the velocity offsets between galaxies and \ion{H}{i} absorbers found near to their major axes with the stellar rotation of the galaxies determined from emission lines in the MUSE data.

We use Astropy \citep{astropycollaboration2018} to fit Gaussian line profiles to each spaxel within the aperture defined by SExtractor. In the case of [\ion{O}{ii}], we fit a double Gaussian with equal line width and redshift-dependent difference in central wavelength, and an intensity ratio between 0.1 and 10 to avoid fitting unphysically large or small values. We use kinematic maps from [\ion{O}{ii}] 3727, [\ion{O}{iii}] 5007 and H$\alpha$ 6563, and generally consider the strongest line available for each galaxy.

\begin{figure}
\includegraphics[width=\columnwidth]{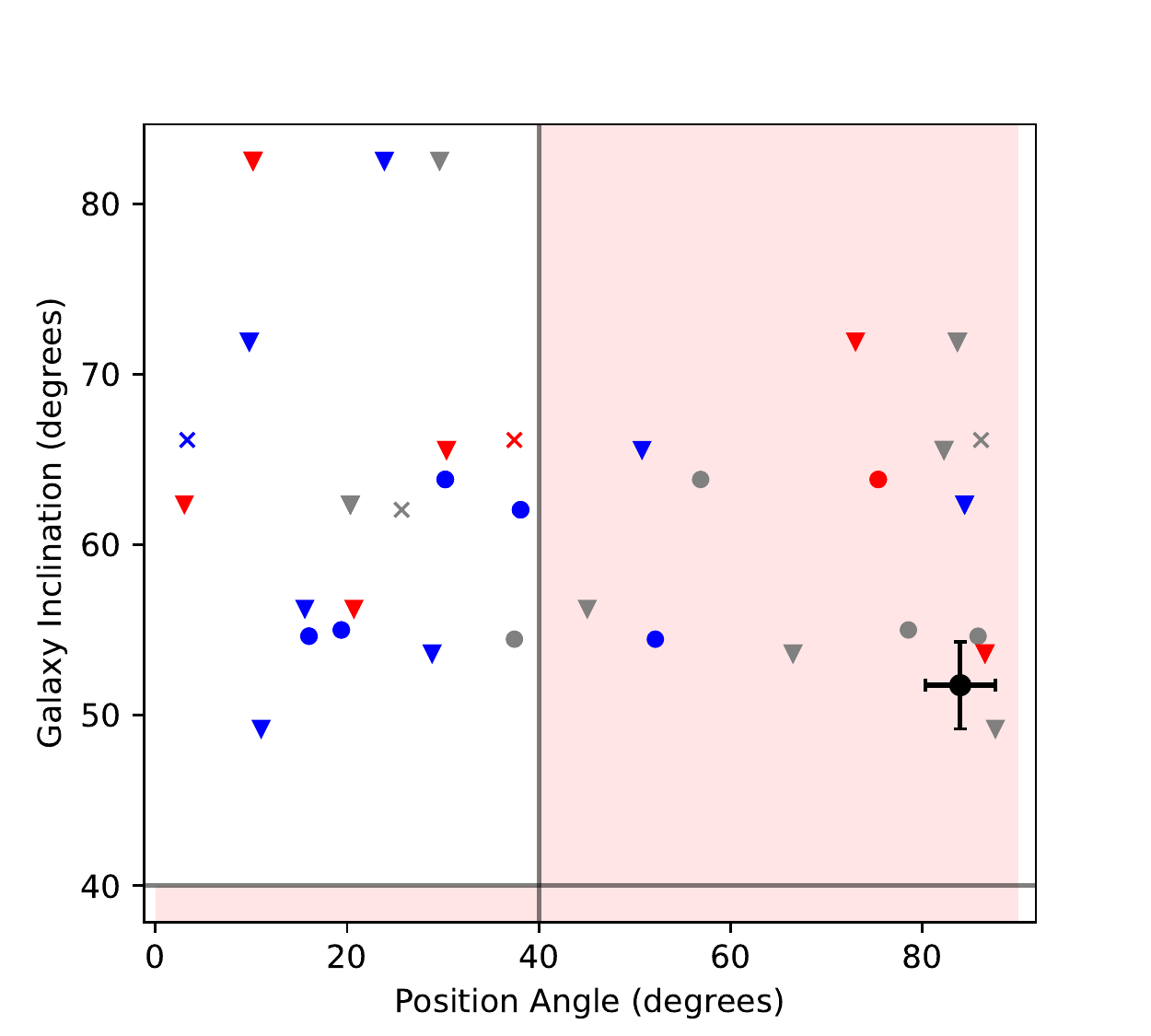}
\caption{The position angles and inclinations of absorption features around galaxies with observed velocity gradients in MUSE (using any emission line). This selection introduces the strong bias against face-on galaxies seen in the figure. We cut by inclination and position angle at $40^{\circ}$ as described in the text, in order to identify candidate co-rotating material (unshaded region). Points are coloured by quasar sightline (A in blue, B in grey, C in red). Galaxies in groups are marked with triangles and non-detections are marked with crosses. The bars on the black point in the lower-right are illustrative of the median 1-$\sigma$ uncertainty in each axis. \label{fig:inc_imp_kinematics}}
\end{figure}
\begin{figure}
\includegraphics[width=\columnwidth]{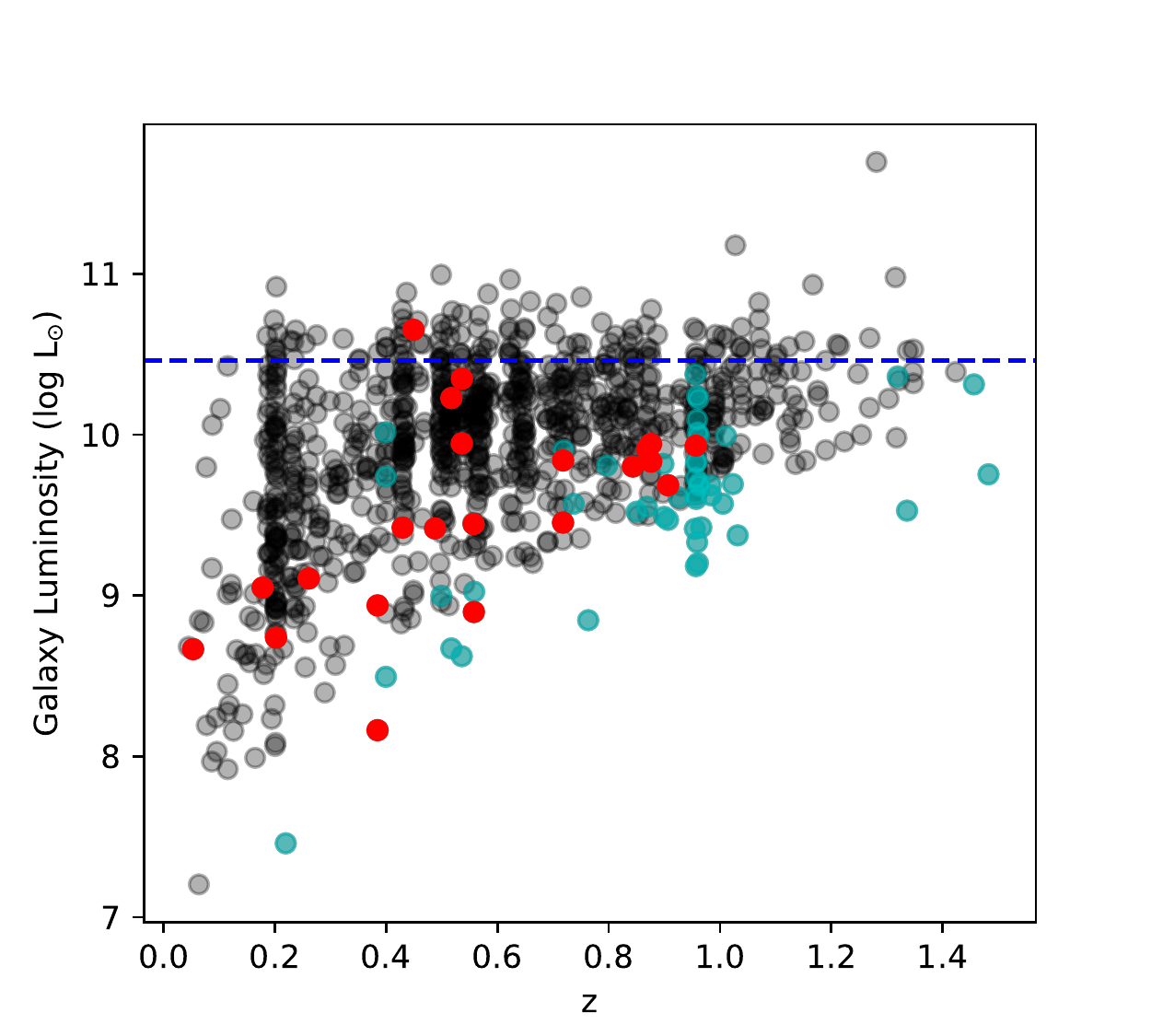}
\caption{The r-band luminosities of observed galaxies as a function of redshift. Red points illustrate objects in which a velocity gradient is observed in MUSE. Cyan points show other galaxies within the MUSE fields. The horizontal dashed line represents an estimate of $L_{\star}$, as given by \citet{montero-dorta2009} for $z\lesssim 0.2.$ ($L_{\star}$ does vary over the given redshift range, but this is expected to be less than a factor of two, e.g. \citealt{gabasch2006}, so is not shown.) \label{fig:lum_z}}
\end{figure}

We consider galaxies in which a velocity gradient is visible in the MUSE data. For comparison with \citet{zabl2019} we only consider galaxies with inclinations above $40^{\circ}$, and position angles on the sky below $40^{\circ}$, although all galaxies with observed velocity gradients already fit this inclination constraint. A velocity window of 500 $\textrm{km}$ $\textrm{s}^{-1}$ is used, as we showed in Section \ref{sec:base} that this includes most of the associated absorption whilst reducing the noise from unassociated IGM absorbers. The position angles and inclinations of the resulting subsample are shown in Figure \ref{fig:inc_imp_kinematics}, whilst the locations of the galaxies in redshift-luminosity space are shown in Figure \ref{fig:lum_z}.

Using the position angle found from GALFIT as described in Section \ref{sec:HST}, we divide the galaxy into two regions, one each side of the projected minor axis. The median velocity of spaxels within each region then gives a redshifted and a blueshifted region. We then compare this to the velocity difference between the galaxy and absorber to identify possible co-rotation. As the subsample is small (22 galaxies with velocity gradients, of which 15 feature major-axis absorption), we also confirm by eye that this algorithm produces the correct result.

\begin{figure}
\includegraphics[width=\columnwidth]{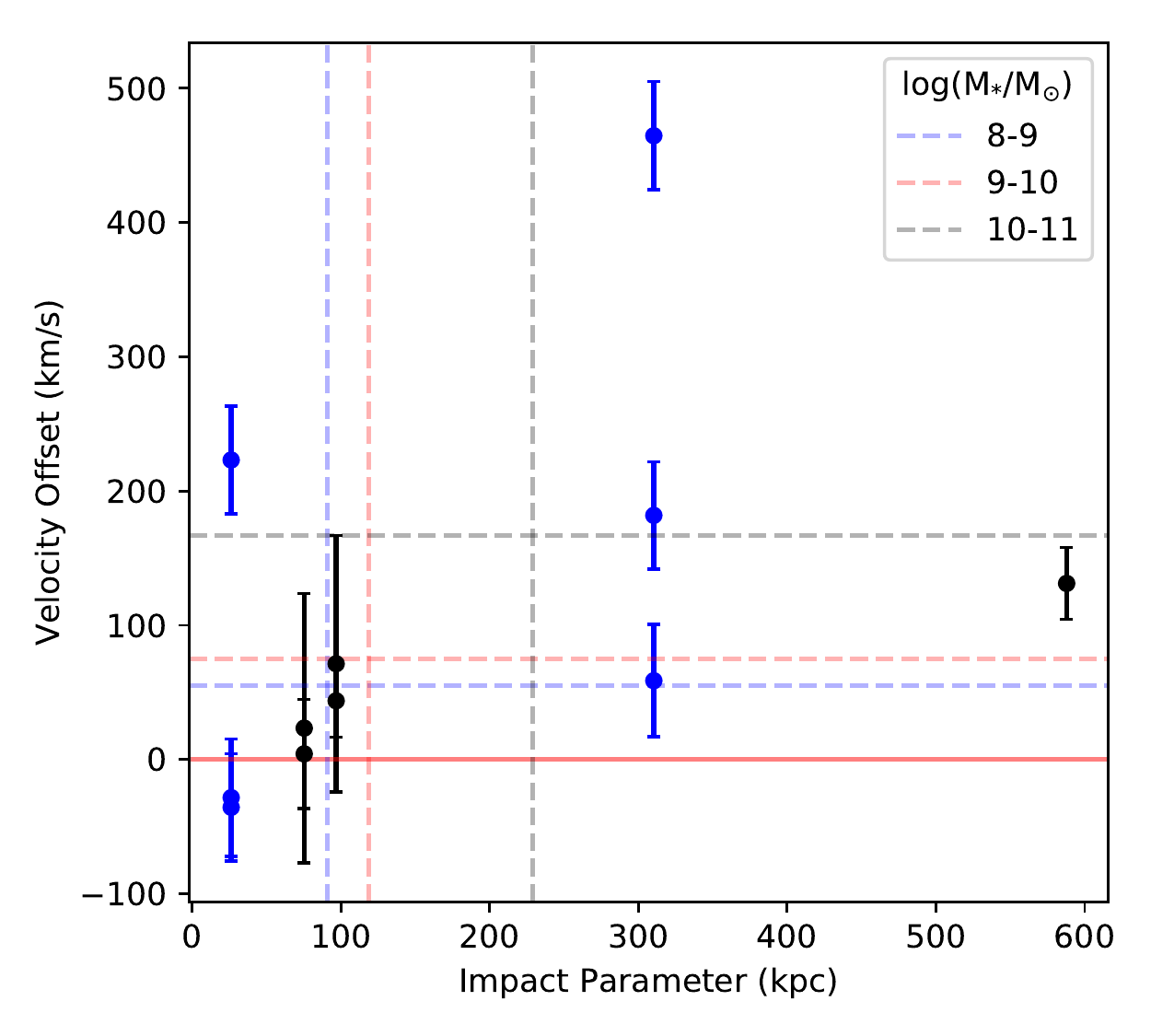}
\caption{The impact parameter and line-of-sight velocity difference between galaxies and absorbers in the major-axis subsample described in Section \ref{sec:kinematics}, showing only non-group galaxies. The sign of the velocity difference is chosen to reflect alignment (positive) between the angular momenta of the galaxy and gas, with the red \added[id=AAB]{solid} line indicating zero offset. Uncertainties in velocity offset are the redshift measurement errors for the galaxy and absorber added in quadrature. Points are coloured by 
\added[id=AAB]{galaxy stellar mass, with galaxies of between $10^{8}$ and $10^{9}$ $\textrm{M}_{\odot}$ shown in blue, $10^{9}$-$10^{10}$ $\textrm{M}_{\odot}$ in red, and $10^{10}$-$10^{11}$ $\textrm{M}_{\odot}$ in black. The dashed horizontal and vertical lines show the median virial velocities and radii for galaxies within each bin across our full galaxy survey}. \label{fig:corotation_results}}
\end{figure}

We first consider only galaxies not lying within groups of five or more galaxies, as any co-rotating structures are less likely to be disturbed by galaxy interactions in these cases. The magnitude and direction of the velocity offset relative to the rotation of the galaxy for each system is illustrated in Figure \ref{fig:corotation_results}. All five of the \added[id=AAB]{galaxy--absorber pairs} meeting the above constraints and not within groups show absorption offset in the same direction as the galaxy rotation. The probability of all five absorbers that are not consistent with zero (within 1$\sigma$) being offset in the same direction if the direction is random is 3\%. A weighted mean of the velocity differences also gives a significant result (105 $\pm$ 40 $\textrm{km}$ $\textrm{s}^{-1}$). \added[id=AAB]{We note that where multiple absorption components are observed in the same spectrum within 500 $\textrm{km}$ $\textrm{s}^{-1}$ of one of these galaxies, each component is counted separately and shown separately in the figures.}

\begin{figure}
\includegraphics[width=\columnwidth]{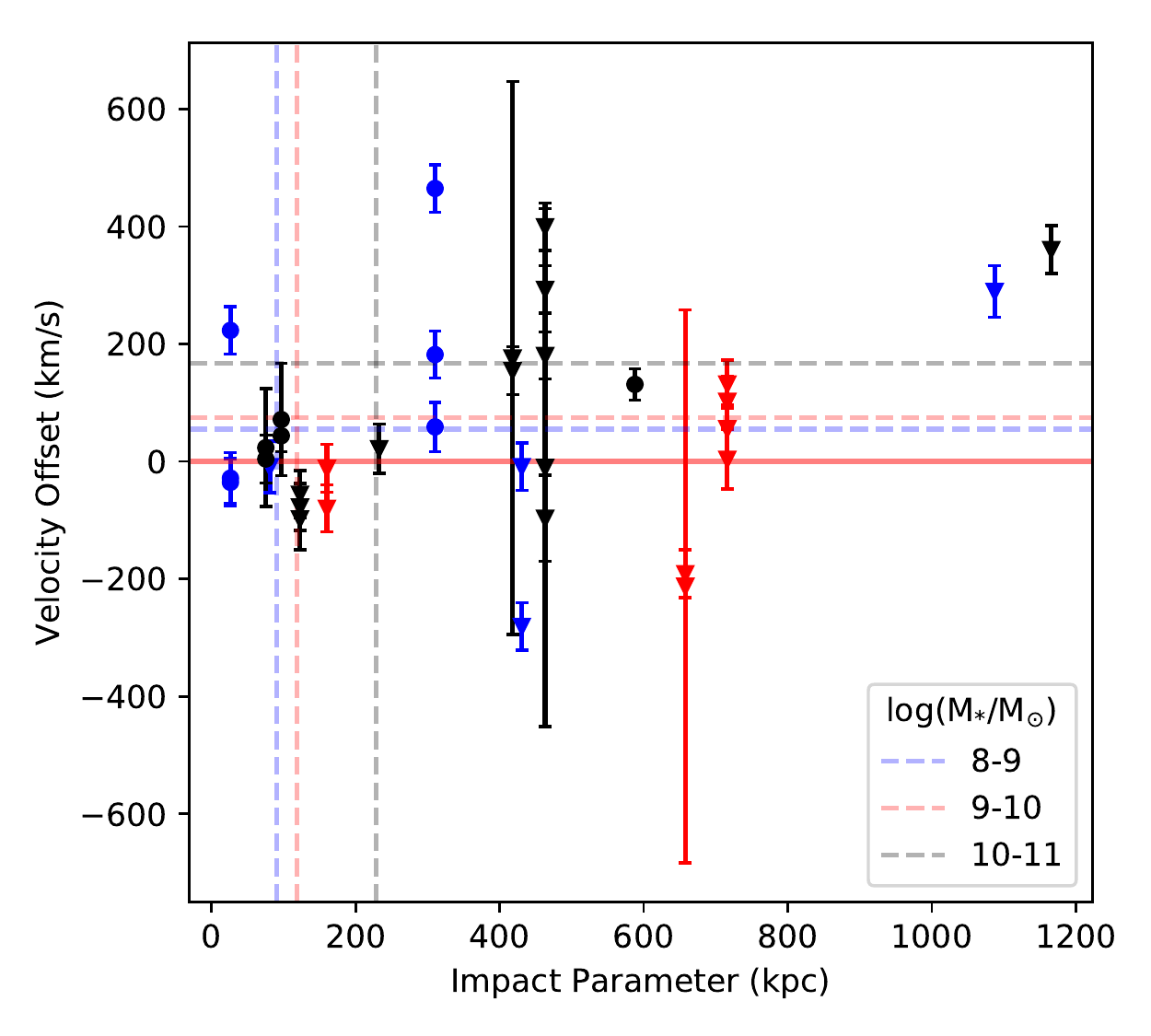}
\caption{The impact parameter and line-of-sight velocity difference between galaxies and absorbers in the major-axis subsample described in Section \ref{sec:kinematics}, as in Figure \ref{fig:corotation_results} but including group galaxies. Galaxies in groups are shown with triangles, those not in groups are shown with circles. Points are coloured
\added[id=AAB]{by galaxy stellar mass. Median virial velocities and radii do not consider other galaxies in the group that may be sharing the same halo}. \label{fig:corotation_results_groups}}
\end{figure}

When galaxies in groups are also included, the major-axis \ion{H}{i} absorption still shows a tendency to align with the galaxy rotation when using a weighted mean (Figure \ref{fig:corotation_results_groups}), with a $\approx$2.5$\sigma$ result of (74 $\pm$ 27) $\textrm{km}$ $\textrm{s}^{-1}$. The binomial calculation returns a p-value of 0.3 (the probability that at least 8 of the 13 absorber velocities not consistent with zero would be aligned with the galaxy rotation if there were no physical link between the galaxy and absorber). 

Whilst the tendency for line-of-sight velocities of gas along the major axis to be aligned with the galaxy rotation would be expected given the rotating disk model, the sample size is too small to draw strong conclusions. Both of the non-group sightlines more than 100 kpc from the galaxy exhibit this alignment well beyond the virial radius of the galaxy, whereas studies such as MEGAFLOW \citep{zabl2019} and others \citep{bouche2016, diamond-stanic2016, ho2017} only show co-rotation out to tens of kpc. In our small sample this can be explained by coincidence, but such alignment is thought to be a result of coherent accretion over Gyr timescales, due to the location of the galaxy within the cosmic web, transferring angular momentum from the gas to the central galaxy \citep[e.g][]{danovich2015, stewart2017}. \citet{defelippis2020} find, in Illustris TNG simulations, that this major-axis co-rotating material often extends in a much thicker `wedge' out to $\sim$ 0.75 $R_{\textrm{vir}}$ for sub-$L_{\star}$ galaxies. However, \citet{huscher2021} use the EAGLE simulations and study larger $\approx L_{\star}$ galaxies, finding substantial rotation only to $\approx 0.3 R_{\textrm{vir}}$ in cool gas at $z=0$. Any attempts to constrain the extent of this material through observations will require detections of absorption with impact parameters of 100--300 kpc from isolated galaxies, which our sample lacks. The sample of \citet{french2020} does contain \ion{H}{i} absorption in this range, but they do not detect a significant preference for co-rotation beyond $\approx$ 100 kpc.


The less-clear alignment in galaxy groups is consistent with models suggesting interactions between galaxies are the origin of most strong \ion{H}{i} absorbers in these environments \citep[e.g][]{morris1994}, as the velocities of such tidal material would be more strongly affected by the relative velocities of the galaxies at large impact parameters. The velocity offsets between galaxies and absorbers are generally within the velocity dispersion of the observed galaxies.

\section{Summary \& Conclusions}\label{sec:conclusions}

In this study we take advantage of multiple observing campaigns, resulting in a deep, dense galaxy survey covering the field around a unique quasar triplet, to examine the geometry and extent of gas flows in the CGM and IGM on larger scales than most CGM studies.

We find that:

\begin{enumerate}
    \item The \ion{H}{i} absorbing gas is often found in structures several hundred kpc in extent, as observed in absorption lines visible in multiple lines-of-sight at the same redshift. These structures traced by \ion{H}{i} over all column densities are more likely to match those traced by galaxies and high-column-density absorbers (N $> 10^{14}$ $\textrm{cm}^{-2}$). We do not find significant evidence that \ion{H}{i} absorbers with lower column densities form structures reaching this extent. It is not clear whether these are arising from smaller overdensities in galaxy voids, or forming filaments, most of which are not aligned such that they cover multiple sightlines. A larger number of background sources would be needed to confirm this through absorption measurements.
    
    \item Galaxy--absorber pairs with \ion{H}{i} absorption exhibit a significant bimodality in position angle for impact parameters up to $\approx$ 300 kpc, possibly evidence that outflows and extended disks reaching these scales are common. However, we do not find that \ion{H}{i} absorbers near the minor axis have significantly higher Doppler widths than those on the major axis, nor do they show smaller velocity offsets around edge-on galaxies for which a putative outflow should not have a large line-of-sight velocity component. We therefore cannot rule out alternative hypotheses for this bimodality, such as preferential alignment between galaxies and large-scale structure. 
    
    \item \ion{H}{i} absorbers found close to the minor axis of a galaxy are twice as likely to have associated \ion{O}{vi} absorption as those found close to the major axis. The position angle distribution of \ion{O}{vi} shows a clear bimodality, weighted towards the minor axis more than \ion{H}{i}, supporting models in which \ion{O}{vi} is primarily observed in galactic winds. \ion{O}{vi} around isolated galaxies is far more limited in extent than \ion{H}{i}, with no isolated galaxies featuring \ion{O}{vi} absorption beyond 400 kpc. We also see \ion{O}{vi} in galaxy groups, but cannot currently distinguish between virialized halo gas, tidal debris, and outflows/halo gas from other galaxies in the same group.
     
    \item The line-of-sight velocities of \ion{H}{i} absorption found near the major axis of a galaxy show a significant tendency to align with the galaxy rotation, suggesting a strong coupling between the angular momentum of the CGM and galaxy \citep[as found by e.g.][]{defelippis2020, huscher2021}.
    
    \item Star-forming galaxies are more likely to trace the same structures as \ion{H}{i} seen in multiple sightlines. Our star-forming sample does not exhibit the bimodality in position angle that could be indicative of accreting and outflowing material, but this bimodality is apparent when using a more restrictive `strongly-star-forming' sample. This suggests that this inflow/outflow dichotomy contributes to the overall bimodality seen on scales $\lesssim$ 300 kpc, although is not ubiquitous among star-forming galaxies, and that the excess absorption seen on larger scales is primarily due to environmental effects (although the redshift biases of our star-forming and non-star-forming samples may also contribute). 
    
    
    
\end{enumerate}


This study illustrates some of the ways in which multiple sightlines can be used to probe the structure of the gas around galaxies. In future work we will also study the absorption around individual galaxies and galaxy groups probed by these quasar lines-of-sight, \added[id=AAB]{where knowledge of absorption at multiple impact parameters and position angles around the same galaxy will enable us to test a range of models}. Similar methods applied to mock lines-of-sight through simulations may provide further insights into the processes behind these results. 

Whilst Q0107 is one of very few systems for which these techniques can currently be used, the prospects at higher redshift are much improved. The ELT will allow far deeper galaxy surveys, enabling high-redshift galaxies to be used as background sources \citep[e.g.][]{japelj2019}, and improving on current studies that have poor signal-to-noise \citep[e.g][]{lee2018} or stack observations to improve the sensitivity \citep[e.g][]{chen2020}. This progresses the field towards full tomography, in which numerous sightlines can probe a variety of scales throughout the same structures, and improve upon the usual pencil-beam observations of the IGM.  

Similarly, observations of emission in \ion{H}{i} 21cm rarely reach detections of material in the $10^{13}-10^{16}$ $\textrm{cm}^{-2}$ regime covered here, but the Square Kilometre Array will likely be able to reach sensitivities of $10^{15}$ $\textrm{cm}^{-2}$ \hbox{\citep{popping2015}}, sufficient to begin detecting the denser filaments of the cosmic web, and allowing study of the interaction between the CGM galactic fountains, the tidal material produced in galaxy groups, and the large-scale cosmic web.

\section*{Acknowledgements}

We thank the referee for useful comments that have improved the quality of this paper.

This work is based on observations made with the NASA/ESA Hubble Space Telescope, obtained from the data archive at the Space Telescope Science Institute. STScI is operated by the Association of Universities for Research in Astronomy, Inc. under NASA contract NAS 5-26555. We also make use of observations collected at the European Southern Observatory under ESO programmes 086.A-0970, 087.A-0857 and 094.A-0131; at the W.M. Keck Observatory under programme A290D; and at the Gemini Observatory under programme GS-2008B-Q-50.

We thank Matteo Fossati for providing the MARZ templates used for estimating redshifts and their uncertainties. We also thank Jill Bechtold for leading the effort to obtain the Keck/DEIMOS data.

We thank the contributors to  SCIPY\footnote{\url{http://www.scipy.org/}}, MATPLOTLIB\footnote{\url{https://matplotlib.org/}}, ASTROPY\footnote{\url{https://www.astropy.org/}}, and the PYTHON programming language, the free and open-source community and the NASA Astrophysics Data system\footnote{\url{https://ui.adsabs.harvard.edu/}} for software and services.

This work also made use of the DiRAC system at Durham University, operated by the Institute for Computational Cosmology on behalf of the STFC DiRAC HPC Facility\footnote{\url{http://www.dirac.ac.uk/}}.

AB acknowledges the support of a UK Science and Technology Facilities Council (STFC) PhD studentship through grant ST/S505365/1. SLM acknowledges the support of STFC grant ST/T000244/1.

SC gratefully acknowledges support from Swiss National Science Foundation grants PP00P2\_163824 and PP00P2\_190092, and from the European Research Council (ERC) under the European Union's Horizon 2020 research and innovation programme grant agreement No 864361. MF also acknowledges funding from the ERC under the Horizon 2020 programme (grant agreement No 757535). This work has been supported by Fondazione Cariplo, grant No 2018-2329.

\section*{Data Availability}

The raw data from the Hubble Space Telescope may be accessed from the MAST archive\footnote{\url{http://archive.stsci.edu/}}, and that from the WM Keck Observatory from the Keck Observatory Archive\footnote{\url{https://koa.ipac.caltech.edu/cgi-bin/KOA/nph-KOAlogin}}. That from the European Southern Observatory may be accessed from the ESO Archive\footnote{\url{http://archive.eso.org/eso/eso_archive_main.html}}, and that from the Gemini Observatory may be accessed from the Gemini Science Archive\footnote{\url{http://www.cadc-ccda.hia-iha.nrc-cnrc.gc.ca/en/gsa/}}. The relevant program IDs are given in Tables \ref{table:qso_summary} and \ref{table:mos_summary} of this paper. The derived data generated in this research will be shared on reasonable request to the corresponding author.

\addcontentsline{toc}{section}{Acknowledgements}




\bibliographystyle{mnras}
\bibliography{zotero_1_Jun_21}



\appendix

\section{Friends-of-Friends Algorithm}\label{sec:fof_groups}

In this section we describe the friends-of-friends algorithm used to find galaxy groups, and the calculations required to determine the linking lengths used. We use the analysis and results from \citet{duarte2014} to motivate our choices.

For physical linking lengths $D_{\perp}$ and $D_{\parallel}$, any pair of galaxies at redshifts $z_{1}$ and $z_{2}$ (with $d_{c}$ the comoving distance to that redshift), separated by an angle $\theta$ on the sky, are `friends' if both of the following conditions are satisfied:

\begin{equation}
    \frac{d_{c}(z_{1}) + d_{c}(z_{2})}{2} \: \theta \: \leq \: D_{\perp}
\end{equation}

\begin{equation}
    \mid d_{c}(z_{2}) - d_{c}(z_{1}) \mid \: \leq \: D_{\parallel}
\end{equation}

We scale the constant linking lengths -- taken from \citet{duarte2014} and denoted b -- by observed galaxy density, $D = b \, n^{-\frac{1}{3}}$, in order to account for the changing magnitude limit with redshift. In order to ensure that large groups do not bias the estimated density, we use large bins with a redshift width 0.2. Note that the angular diameter distance could be used in place of the comoving distance, and it could be argued that $d_{A}$ should be used since galaxy groups are expected to be virialized. However, this also requires the line-of-sight linking length to be reduced by a factor of (1+z), and most studies do not apply this.

The comoving volume is calculated by integrating a conic section with an half-opening angle of $0.15^{\circ}$, through comoving distance between the redshift limits. This is approximately the radius of the DEIMOS data. MUSE galaxies are not counted in this process, as the greater depth of the MUSE data would bias the selection of groups in and around the MUSE fields (by adding galaxies to groups that would be undetected if they lay outside of the MUSE fields, and potentially joining groups together that would otherwise remain separate). By excluding these galaxies, the depth of VIMOS, DEIMOS and GMOS coverage throughout this aperture is close to constant.

\begin{equation}
    V = \int_{d_{c}(z_{min})}^{d_{c}(z_{max})} \pi (\theta d_{c})^{2} \: d(d_{c}) 
\end{equation}

In each redshift bin, the number of observed galaxies is divided by the volume of the cone, to give the galaxy number density, which is converted to a linking length as above. Whilst the physical linking length is constant within a single bin, the angular separation and redshift difference are both affected by the galaxy redshift itself.

This algorithm is implemented by assigning each galaxy a group identifier. We then loop through each galaxy pair, and assign both galaxies the lower identifier if they are `friends'. This is repeated until the sum of all identifiers converges (usually 3 to 5 iterations). All galaxies with the same identifier form a single galaxy group. Absorbers and the MUSE galaxies are associated with galaxy groups if they are 'friends' with any galaxy in that group. 

\citet{duarte2014} list three pairs of linking lengths, from which we use the larger, completeness-optimizing, values ($b_{\perp} = 0.2$, $b_{\parallel} = 3$). This results in linking velocity differences of $\approx 800-1500$ km/s and transverse linking lengths of $\approx 800-1200$ pkpc. These linking lengths are large enough to allow a reasonable sample size of galaxy groups, as well as approximately matching the distance scales separating the three lines-of-sight. Such large linking lengths ensure that our non-group sample excludes any galaxies whose CGM is likely to be affected by ongoing group interactions.

We do repeat most analyses shown in the main text whilst using the `medium' linking lengths from \citet{duarte2014}, and this does not reveal any significant differences. Using the shortest linking lengths, we do not find enough groups to draw any conclusions.

\section{Generating Random Absorbers} \label{sec:random_absorbers}

In Section \ref{sec:coherence}, we test the excess of galaxy-absorber associations within a series of different constraints. Here we describe the process of generating the random sets of absorbers used in these tests, in which any physical association between gas and galaxies is removed. 

We generated 5000 sets of randomly distributed absorbers similar to the method used in T14, as follows:

\begin{enumerate}
 \item Calculate the signal-to-noise per resolution element for QSO spectrum
 \item Convert this to minimum rest-frame equivalent width for the absorption feature as a function of redshift (see Figure \ref{fig:ew_detection})
 \item For each real absorption feature, find the allowed region in redshift space for which the EW of the progenitor is larger than the minimum, and is not covered by galactic absorption
\item Distribute absorbers randomly through the allowed region,  giving the random absorber the same properties as the observed progenitor
\end{enumerate}

The signal-to-noise per resolution element is estimated by dividing the continuum value of the QSO spectrum by the uncertainty in the value of that pixel, and then convolving with the line-spread function of the instrument. In order to improve execution time we use a box car of the correct FWHM as an approximation to the line-spread function.

The minimum observable equivalent width for an unresolved transition observed at wavelength $\lambda$ is:

\begin{equation}
    W_{min}(\lambda) = S \frac{FWHM}{\langle SNR \rangle_{\lambda}}
\end{equation}

\noindent
where FWHM is the width of the line-spread of the instrument, S is the significance required of the detection in units of sigma (where a value of 3 best matches the distribution of observed absorbers), and $\langle SNR \rangle_{\lambda}$ is the mean signal-to-noise per resolution element. If the resolution R = $\frac{\lambda}{FWHM}$ is constant:

\begin{equation}
    W_{min}(\lambda) = \frac{S \lambda}{R \langle SNR \rangle_{\lambda}} = \frac{S \lambda_{0}(1+z)}{R \langle SNR \rangle_{\lambda}}
\end{equation}

\noindent
where $\lambda_{0}$ is the rest-frame wavelength of the transition. This gives a rest-frame equivalent width:

\begin{equation}
    W_{min, r}(z) = \frac{S \lambda_{0}}{R \langle SNR \rangle_{z}}
\end{equation}

\added[id=AAB]{We note that the lines in the COS spectra are usually resolved, with a median width $\textrm{b} \approx 30 \textrm{km}$ $\textrm{s}^{-1}$. For these lines a more representative equivalent width limit can be found using the prescription given in \citet{keeney2012}, in which the optimal integration window is found using a convolution of the COS line-spread-function and the absorption line profile, rather than just the resolution of the instrument. However, many of the absorption features are narrower than this, so we continue to use the COS FWHM as a limiting value. This only has a notable effect on absorbers that are both weak and narrow, so only four absorbers of the 184 across the three COS spectra would have their allowed redshift ranges reduced by more than 15\% if the \citet{keeney2012} prescription were used.}

From the absorption line catalogue (taken directly from T14), the best fit column density and Doppler parameter are given for each absorber. This can be robustly converted to an equivalent width via optical depth using the approximation given in \citet{draine2011} (eqn 9.8, assuming negligible stimulated emission):

\begin{equation}
    \tau_{0} = \sqrt{\pi} \frac{e^2}{m_{e} c} \frac{n f_{osc} \lambda_{0}}{b}
\end{equation}

\noindent
where n and b are the column density and Doppler parameter of the absorber, $\lambda_{0}$ the rest-frame wavelength, and $f_{osc}$ the oscillator strength. Draine then provides an approximation for equivalent width:

\begin{equation}
    W = \sqrt{\pi} \frac{b}{c} \frac{\tau_{0}}{1+(\tau_{0}/2\sqrt{2})} \textrm{(for $\tau_{0} <$  1.25393)}
\end{equation}

\begin{equation}
    W = \sqrt{\frac{4b^2}{c^2} ln(\frac{\tau_{0}}{ln 2}) + \frac{b}{c} \frac{\gamma \lambda_{0}}{c} \frac{\tau_{0} - 1.25393}{\sqrt{\pi}}} \textrm{(for $\tau_{0} >$ 1.25393)}
\end{equation}

\noindent
$\gamma$ is the damping term of the Lorentzian component of the Voigt profile, which becomes non-negligible for high-column-density systems.

This value of equivalent width can then be compared with the minimum equivalent width detectable in the spectrum as a function of redshift, shown for QSO-A in Figure \ref{fig:ew_detection}. We mask out the regions where the absorber would be undetectable due to insufficient equivalent width, as well as regions within 200 km/s of galactic absorption (as in T14, using \ion{C}{ii}, \ion{N}{v}, \ion{O}{i}, \ion{Si}{ii}, \ion{S}{ii} and \ion{Fe}{ii}). The region less than 3500 km/s blueward of the QSO is also masked, in order to remove most associated absorption. As found by \hbox{\citet{wild2008}}, $\approx$ 40\% of \ion{C}{iv} absorption out to 3000 km/s is associated with outflows from the QSO, although absorption from outflows can extend beyond 10,000 km/s. The velocity cut used is therefore a compromise between excluding most absorption associated with the QSO, whilst excluding minimal intervening IGM absorption. This cut is also much larger than the region within which the proximity effect is significant \citep[e.g][]{scott2002}. Absorbers in this set that have already been placed are also masked using a window of 35 km/s (approximately the median width of \ion{H}{i} absorption systems) so that systems that would not be resolved due to blending are not generated by this process. The absorber is then randomly assigned a redshift within the unmasked region.

\section{GALFIT results} \label{sec:galfit_results}

\begin{figure*}
\includegraphics[width=0.98\textwidth]{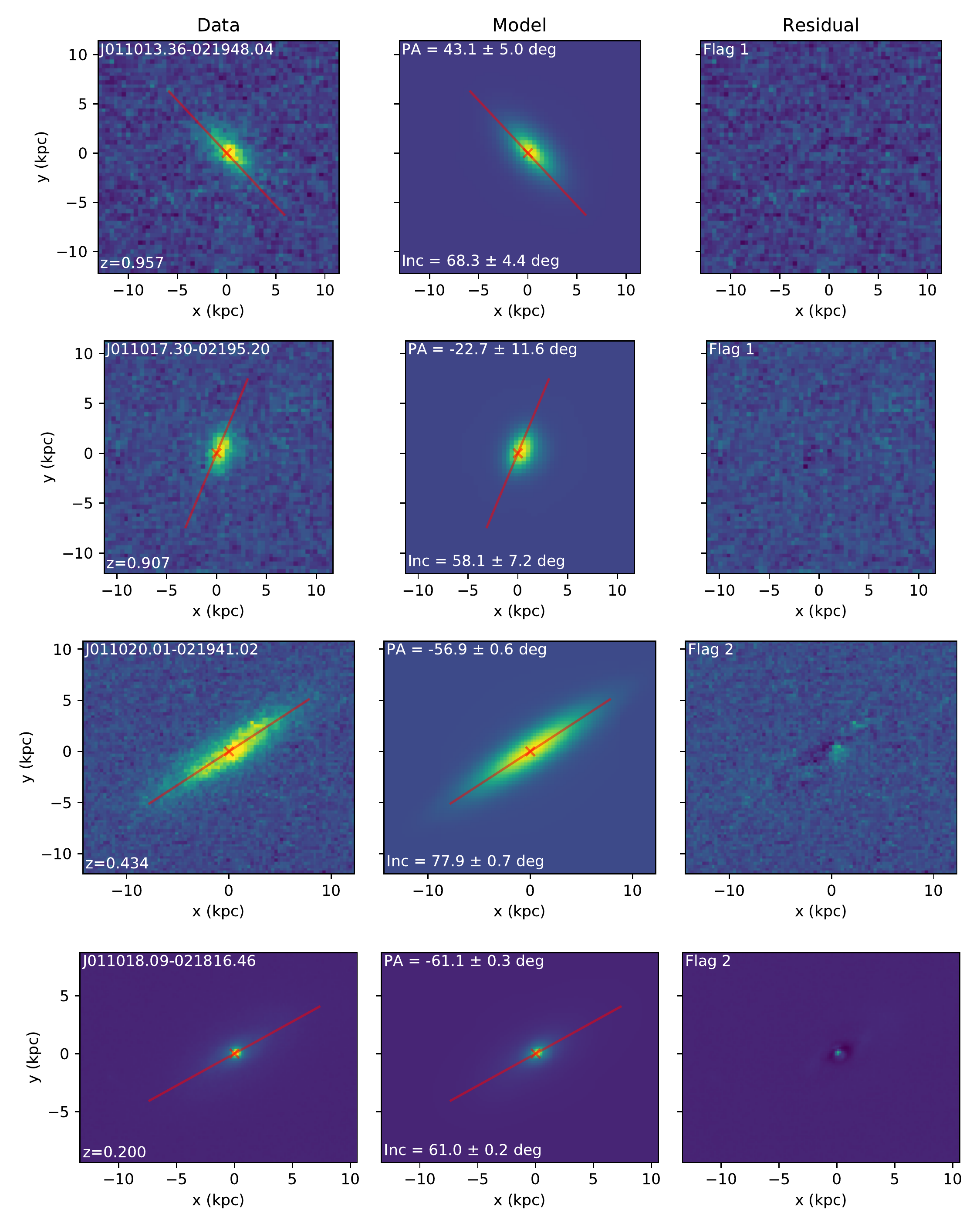}
\caption{\added[id=AAB]{Several examples of galaxy cut-outs resulting from our GALFIT modelling. Each row shows a single galaxy, with all three panels on an identical flux scale. The left panel illustrates the data from HST, the middle panel the best fit GALFIT model, and the right the residual. The galaxy ID and redshift, as well as the resulting position angle (relative to the HST image, which is not aligned with north), inclination and quality flag, are shown in the panels. The projected major axis, as determined by GALFIT, is shown by the red line.} \label{fig:hst_examples_1}}
\end{figure*}

\begin{figure*}
\includegraphics[width=\textwidth]{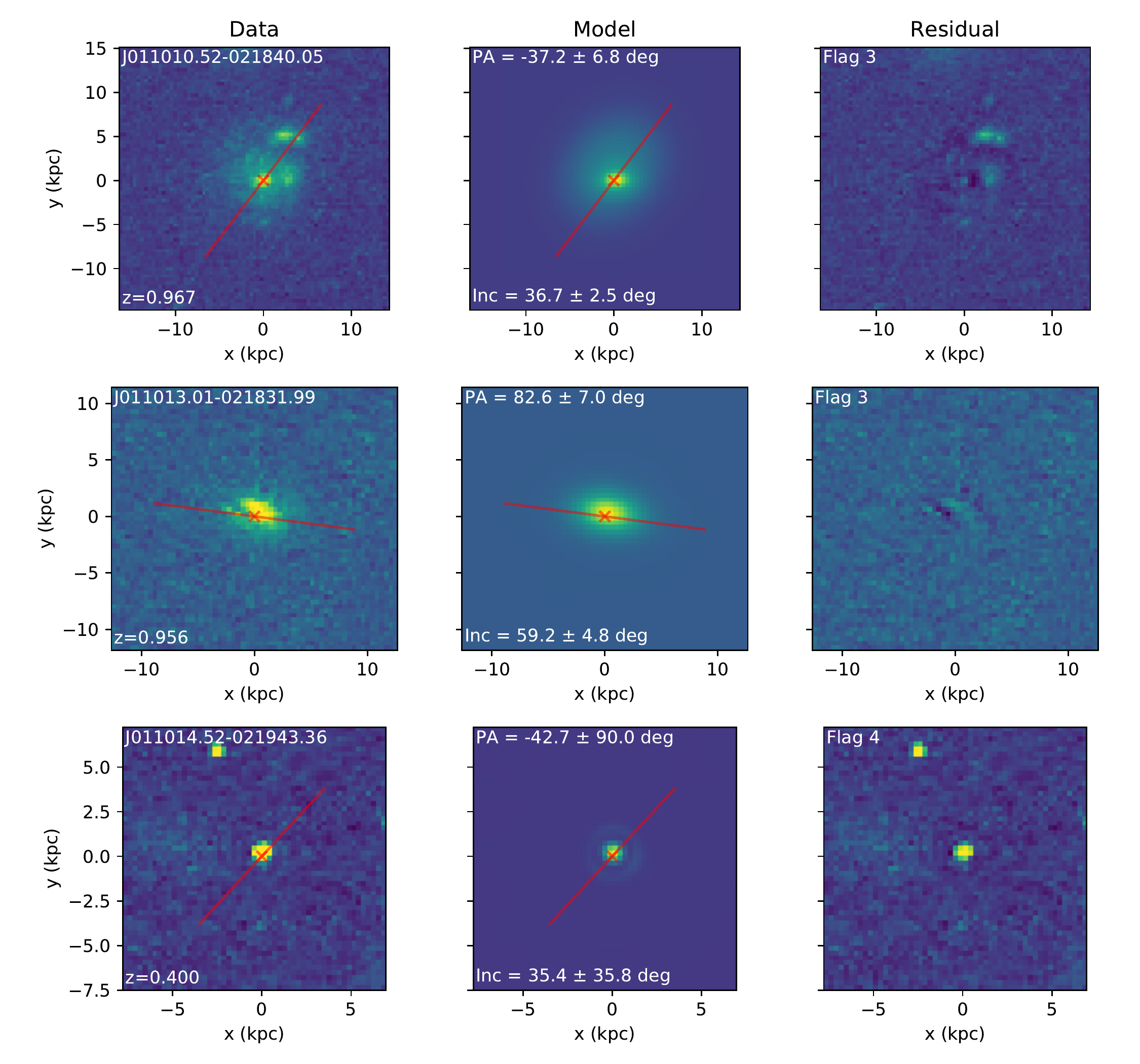}
\caption{\added[id=AAB]{Further examples of galaxy cut-outs resulting from our GALFIT modelling, as in Figure \ref{fig:hst_examples_1}} \label{fig:hst_examples_2}}
\end{figure*}

\added[id=AAB]{Here we include some examples of our GALFIT results, in order to illustrate the utility of the modelling and the use of the quality flags described in Section \ref{sec:HST}. As illustrated in the Figures \ref{fig:hst_examples_1} and \ref{fig:hst_examples_2}, the quality flags are based on the goodness of fit, not on the uncertainty in position angle or inclination. The flag 1 objects have no clear structure remaining in their residuals, flag 2 objects have minor structures, flag 3 objects may have major structures in the residuals that could indicate a poor estimate of position angle and inclination, and flag 4 objects are clearly a poor fit. Our tests in Section \ref{sec:orientations} are run using objects with flags 1-2 and 1-3, with similar results obtained in each case.}

\begin{figure*}
\includegraphics[width=\textwidth]{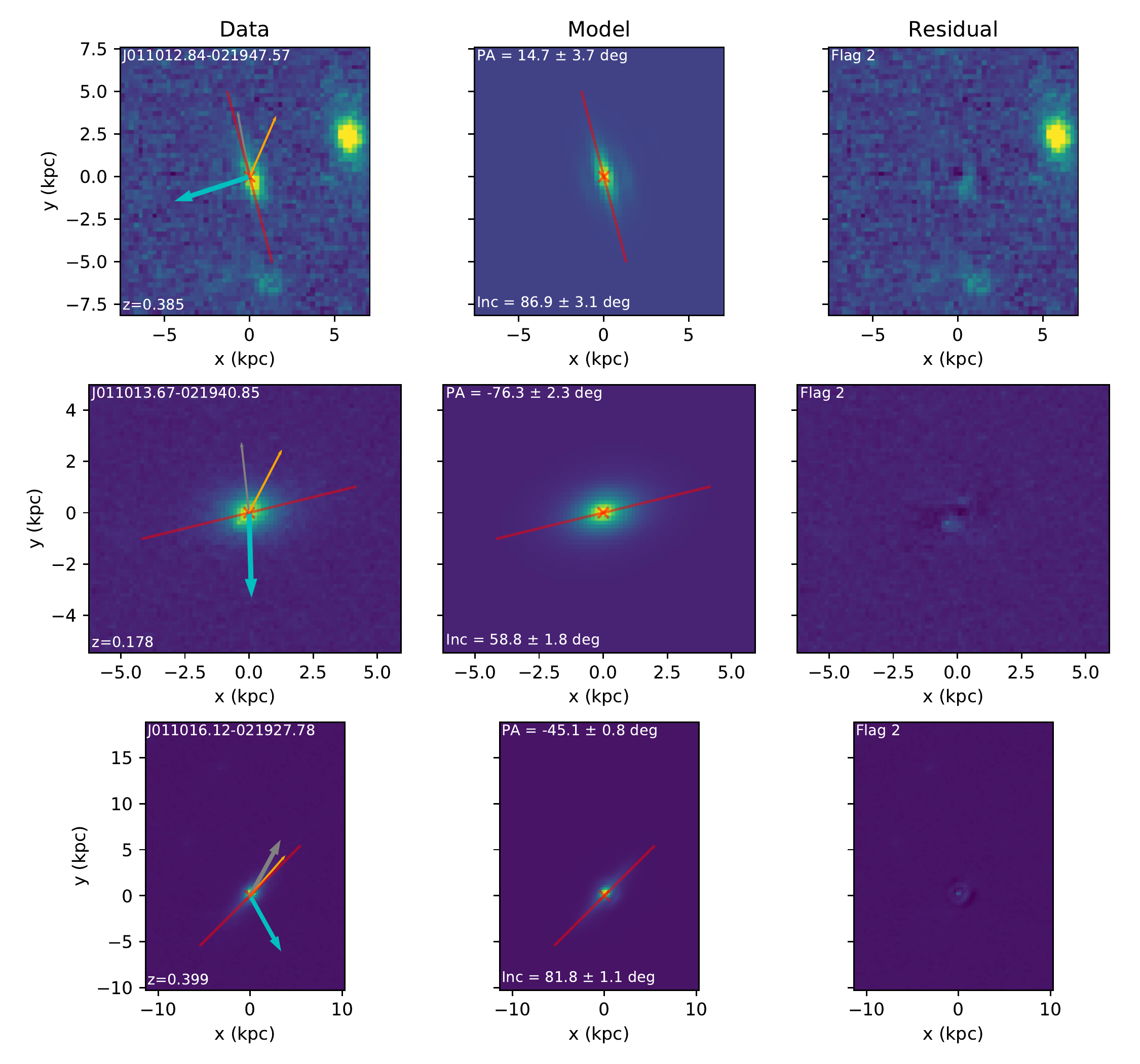}
\caption{\added[id=AAB]{Galaxy cut-outs resulting from our GALFIT modelling, showing galaxies not in groups, but with observed \ion{O}{vi} absorption (as discussed in Section \ref{sec:metals}). Each row shows a single galaxy, with all three panels on an identical flux scale. The left panel illustrates the data from HST, the middle panel the best fit GALFIT model, and the right the residual. The galaxy ID and redshift, as well as the resulting position angle (relative to the HST image, which is not aligned with north), inclination and quality flag, are shown in the panels. The projected major axis, as determined by GALFIT, is shown by the faded red line. The three coloured stubs in the left-hand panel point the direction of the three QSOs, with cyan towards A, grey towards B, and orange towards C. The thick stubs with arrowheads indicate that that sightline exhibits detected \ion{O}{vi} absorption within $500 \textrm{km}$ $\textrm{s}^{-1}$, whilst the thin stubs without arrowheads inidcation non-detections of \ion{O}{vi}.} \label{fig:hst_examples_o6_1}}
\end{figure*}

\begin{figure*}
\includegraphics[width=\textwidth]{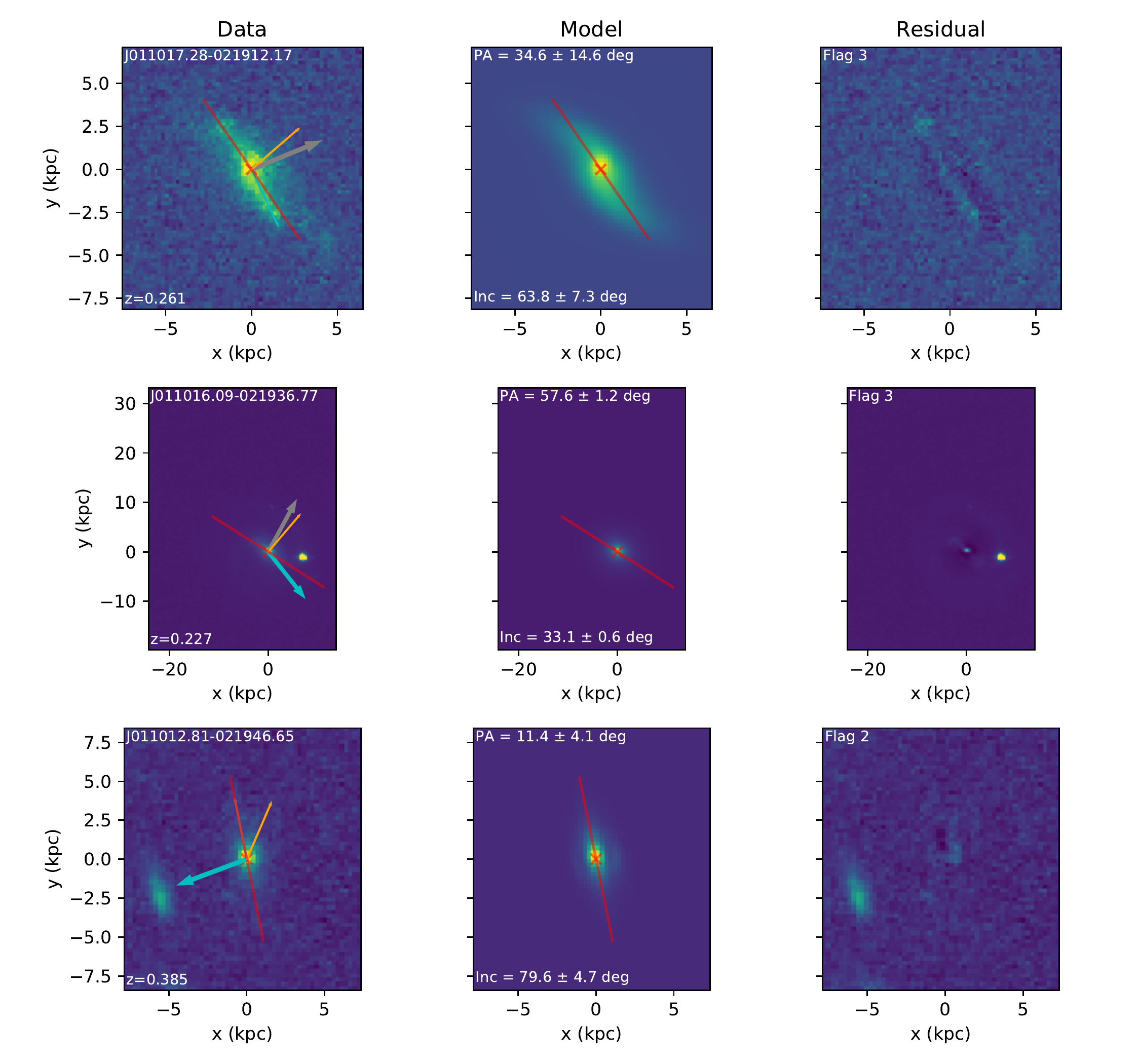}
\caption{\added[id=AAB]{Further examples of galaxy cut-outs of objects with \ion{O}{vi} absorption detected within $500 \textrm{km}$ $\textrm{s}^{-1}$, as in Figure \ref{fig:hst_examples_o6_1}.} \label{fig:hst_examples_o6_2}}
\end{figure*}

\added[id=AAB]{Figures \ref{fig:hst_examples_o6_1} and \ref{fig:hst_examples_o6_2} show all non-group galaxies with reasonable GALFIT results (at least flag 3) that exhibit \ion{O}{vi} absorption within $500 \textrm{km}$ $\textrm{s}^{-1}$. These illustrate the tendency for \ion{O}{vi} absorption to be detected along the minor axis, as discussed in Section \ref{sec:metals}.}

\bsp	
\label{lastpage}
\end{document}